\documentclass[twocolumn]{aastex63}% [manuscript]{aastex63} %
\newcommand{\meth}{CH$_3$OH}
\newcommand{\droh}{DR21(OH)}
\newcommand{\kms}{km~s$^{-1}$}
\newcommand{\jybeam}{Jy~beam$^{-1}$}
\newcommand{\vsp}[1]{\vspace{#1in}}
\accepted{March 2022 for publication in the Astrophysical Journal}
\shorttitle{Long-term variability of Class~I methanol masers}
\shortauthors{Wenner et al.}

\begin{document}

\title{Long-term variability of Class~I methanol masers in the \\ high mass star forming region DR21(OH)}

%% \author[xxxx-xxxx-xxxx-xxxx]{Author Name} % Now takes 16 digit ORCID

\correspondingauthor{A. P. Sarma \hfill asarma@depaul.edu}
%\email{asarma@depaul.edu}

\author{Nycole Wenner}
\affiliation{CIERA, Department of Physics \& Astronomy, Northwestern University, Evanston IL 60208, USA}

\author{A. P. Sarma}
\affiliation{Department of Physics \& Astrophysics, DePaul University \\
2219 N. Kenmore Ave, Byrne 211, Chicago IL 60614, USA}

\author{E. Momjian}
\affiliation{National Radio Astronomy Observatory \\
P. O. Box O, Socorro, NM, USA}

\begin{abstract}

\noindent
High mass stars play an important role in the Interstellar Medium, but much remains to be known about their formation. Class~I methanol masers may be unique tracers of an early stage of high mass star formation, and a better understanding of such masers will allow them to be used as more effective probes of the high mass star forming process. We present an investigation of the long-term variability of Class I methanol masers at 44 GHz toward the high mass star forming region DR21(OH). We compare observations taken in 2017 to 2012 and also to 2001 data from the literature. A total of 57 maser spots were found in the 2017 data, with center velocities ranging between $-8.65$~\kms\ to +2.56~\kms. The masers are arranged in a western and an eastern lobe with two arcs in each lobe that look like bowshocks, consistent with previous observations. The general trend is an increase in intensity from 2001 to 2012, and a decrease from 2012 to 2017. Variability appears to be more prevalent in the inner arc of the western lobe than in the outer arc. We speculate that this may be a consequence of episodic accretion, in which a later accretion event has resulted in ejection of material whose shock reached the inner arc at some point in time after 2001. We conclude that class I methanol masers are variable on long timescales (of the order of 5-10 years).

\vspace{0.5in}

\end{abstract}


\section{Introduction} \label{sec:intro}

Masers are bright and compact sources, and offer the ability to observe high mass star forming regions at high angular resolution \citep{richards+2020}. High mass stars ($M \gtrsim 8~M_\odot$) play an important role in shaping their surroundings and the Interstellar Medium (ISM), by driving strong winds and massive outflows, emitting UV radiation to create expanding ionized regions, and ending in supernova explosions that are the principal source of heavy elements in the ISM \citep{zy+2007}. Yet, observing the formation of high mass stars and establishing an evolutionary sequence remains a challenge because they are rare, located farther away from us, and tend to form in clusters \citep{motte+2018}. Recently, it has also become clear that high mass stars may undergo episodic accretion during their formation, just like in low mass star formation (e.g., \citealt{caratti+2017}; \citealt{brogan+2019}).

\vsp{0.1}

A better understanding of maser characteristics makes them more effective probes of high mass star forming regions. One such observable property is the variability of masers. Indeed, the connection of the variability of so-called Class~II methanol (\meth) masers to changes in the radiative output of high mass protostars due to episodic accretion has already been established (e.g., \citealt{hunter+2018}). The variability of Class~I \meth\ masers, on the other hand, has only been hinted at in the literature (e.g., \citealt{kurtz+2004, momjian+2017}). Class~I \meth\ masers in high mass star forming regions are generally found in outflows, where shocks provide the collisional pumping that causes the population inversion necessary for maser action. They are likely unique tracers of the very early stages of high mass star formation \citep{leurini+2016}. Since each episode of accretion in high mass star formation is accompanied by outward-propagating shocks (\citealt{caratti+2015}), the variability of Class~I~\meth\ masers could potentially be linked to such accretion events.

\vsp{0.05}

In this paper, we present a dedicated investigation of the variability of Class I \meth\ masers at 44 GHz in the high mass star forming region \droh. This region is located in the northern part of the Cygnus-X molecular cloud complex, at a distance of 1.5~kpc \citep{rygl+2012}. The 44~GHz Class~I \meth\ masers in \droh\ are arranged in a low-velocity outflow with a redshifted western lobe and a predominantly blueshifted eastern lobe (\citealt{kogan+1998}; \citealt{kurtz+2004}). In each of these two lobes, the masers are arranged in two arcs \citep{araya+2009}. The arcs traced by the Class~I \meth\ masers correspond to bow shocks created by the outflows.  \citet{orozco+2019} found that the thermal \meth\ emission at millimeter wavelengths follows a morphology very similar to the low-velocity outflow traced by the 44~GHz \meth\ masers, with the emission concentrated into compact bow shock structures along the outflow. Earlier, \citet{zapata+2012} found that these bow-shocked structures are also traced by millimeter thermal emission lines of formaldehyde (H$_2$CO). Observing with the Smithsonian Millimeter Array (SMA), \citet{zapata+2012} found a 1.4 mm source, SMA4, which they proposed as the source of the outflow. In our Figure~\ref{fig:2017map} (introduced in Section~\ref{sec:res}), we have marked the position of SMA4 with a blue ``+'' sign, and indicated the roughly east-west direction of the 44~GHz \meth\ maser outflow with solid red and blue arrows. High-velocity outflows in the CO ($J$ = 2$-$1) line are also observed in \droh, but they appear to arise from a different source than the 44~GHz \meth\ maser outflow (\citealt{zapata+2012}). The source of the CO outflows is about $2\farcs7$ to the east and north of SMA4, but still within the 2.7~mm dust source MM2 in \droh\ that was resolved into four sources with the SMA at 1.4 mm (\citealt{zapata+2012}, and references therein). One of the high-velocity CO ($J$ = 2$-$1) outflows has the same east-west orientation as the \meth\ maser outflow, but its redshifted lobe is to the east and its blueshifted lobe to the west (\citealt{zapata+2012}). In Figure~\ref{fig:2017map}, we have indicated the direction of this CO outflow with dashed blue and red arrows, and marked the source at the origin of the outflow with a purple cross (following Figure 3 of \citealt{zapata+2012}). Another high-velocity CO ($J$ = 2$-$1) outflow is monopolar, with blueshifted emission toward the southwest; it is also marked in Figure~\ref{fig:2017map}. Additional discussion of high-velocity outflows in the CO ($J$ = 3$-$2) line is in \citet{girart+2013}.

\begin{deluxetable}{lcrrrrrrrrcrl}
\tablenum{1}
\tablewidth{0pt}
\tablecaption{Parameters for VLA Observations
	\protect\label{tOP}}
\tablehead{
\colhead{Parameter} & 
\colhead{Value} }
\startdata
Date \dotfill & 2017 May 24    \\
Configuration \dotfill & C \\
R.A.~of field center (J2000) \ldots & 20$^{\text{h}}$~39$^{\text{m}}$~00$\fs$8 \\
Dec.~of field center (J2000) \dotfill & 42\arcdeg~22\arcmin~47\rlap{\arcsec}.\,0\\
Total bandwidth (MHz) \dotfill & 4.0 \\
No.~of channels \dotfill & 1024 \\
Channel spacing (km~s$^{-1}$) \dotfill & 0.0266 \\
Approx.~time on source \dotfill & 81~min \\
Rest frequency (GHz) \dotfill  & 44.069488 \\
FWHM of synthesized beam \dotfill & $ 0\, \rlap{\arcsec}.\, 58 \times 0\, \rlap{\arcsec}.\, 55$ \\
& P.A. = $-$65.88\arcdeg \\ 
Line rms noise (mJy~beam$^{-1}$) \tablenotemark{a} & 14 \\
\enddata
%\tablecomments{}
\tablenotetext{a}{The line rms noise was measured from the Stokes $I$ image cube using maser line free channels.}
\end{deluxetable}

In this paper, we investigate the variability of these Class~I \meth\ masers by comparing data taken in 2017 with observations from 2012 reported in \citet{momjian+2017}, and 2001 data taken from the literature \citep{araya+2009}. This investigation was motivated by the finding in \citet{momjian+2017} that the two strongest 44~GHz Class~I \meth\ masers in \droh\ had swapped rank from 2001 to 2012. What had been the strongest maser in the outer arc of the western lobe in 2001 \citep{araya+2009} became the second strongest maser in 2012, whereas the second strongest maser in the inner arc of the western lobe became the strongest maser. The details of the 2017 observations and data reduction are given in \S~\ref{sec:obsredn}. The results are presented in \S~\ref{sec:res}, and discussed in \S~\ref{sec:disc}. In \S~\ref{sec:conc}, we state our conclusions.

\vsp{0.025}

\section{Observations and Data Reduction} \label{sec:obsredn}
% Useful Info: Proposal code is 17A-038

We observed the $7_{0}-6_1\, A^+$ Class~I \meth\ maser emission line at 44~GHz toward the high mass star forming region \droh\ with the Karl G. Jansky Very Large Array (VLA) of the NRAO\footnote{The National Radio Astronomy Observatory (NRAO) is a facility of the National Science Foundation operated under cooperative agreement by Associated Universities, Inc.} on 2017 May 24. The VLA was in the C-configuration with a maximum baseline length of 3.4\,km. The Wideband Interferometric Digital ARchitecture (WIDAR) correlator of the VLA was configured to deliver a single 4 MHz sub-band with 1024 spectral channels. The resulting channel spacing was 3.90625 kHz, corresponding to 0.0266~\kms\ at the observed frequency. The parameters for our VLA observations are summarized in Table~\ref{tOP}. We included observations of the source J1331+3030 (3C286) to calibrate the absolute flux density scale. Even though the maser emission from DR21(OH) is strong enough for self-calibration, we also observed the nearby compact radio source   J2007+4029 in part of the observing session to calibrate the complex gains. The latter calibrator was observed to derive the absolute positions of the masers in the target source. The cycle time between the target and this nearby calibrator was five minutes.

\vsp{0.05}

Calibration, deconvolution, and imaging were performed using the Astronomical Image Processing System (AIPS). The spectral line data of the target source DR21(OH) were Doppler corrected, and the frequency channel with the brightest maser emission signal was split off and self-calibrated first in phase, then in both phase and amplitude, and imaged in a succession of iterative cycles. The final self-calibration solutions were applied to the full spectral line data set. The final image cube was made with a synthesized beamwidth of $0\, \rlap{\arcsec}.\, 58 \times 0\, \rlap{\arcsec}.\, 55$ (Table~\ref{tOP}). The AIPS tasks TVSAD and ISPEC were then used to search for and verify the presence of masers in the image cube, and the task XGAUS was used to fit Gaussian components to the spectral profiles of the identified masers. 

\vsp{0.025}

\section{Results} \label{sec:res}

\begin{figure}[htb!]
\epsscale{0.7}
\plotone{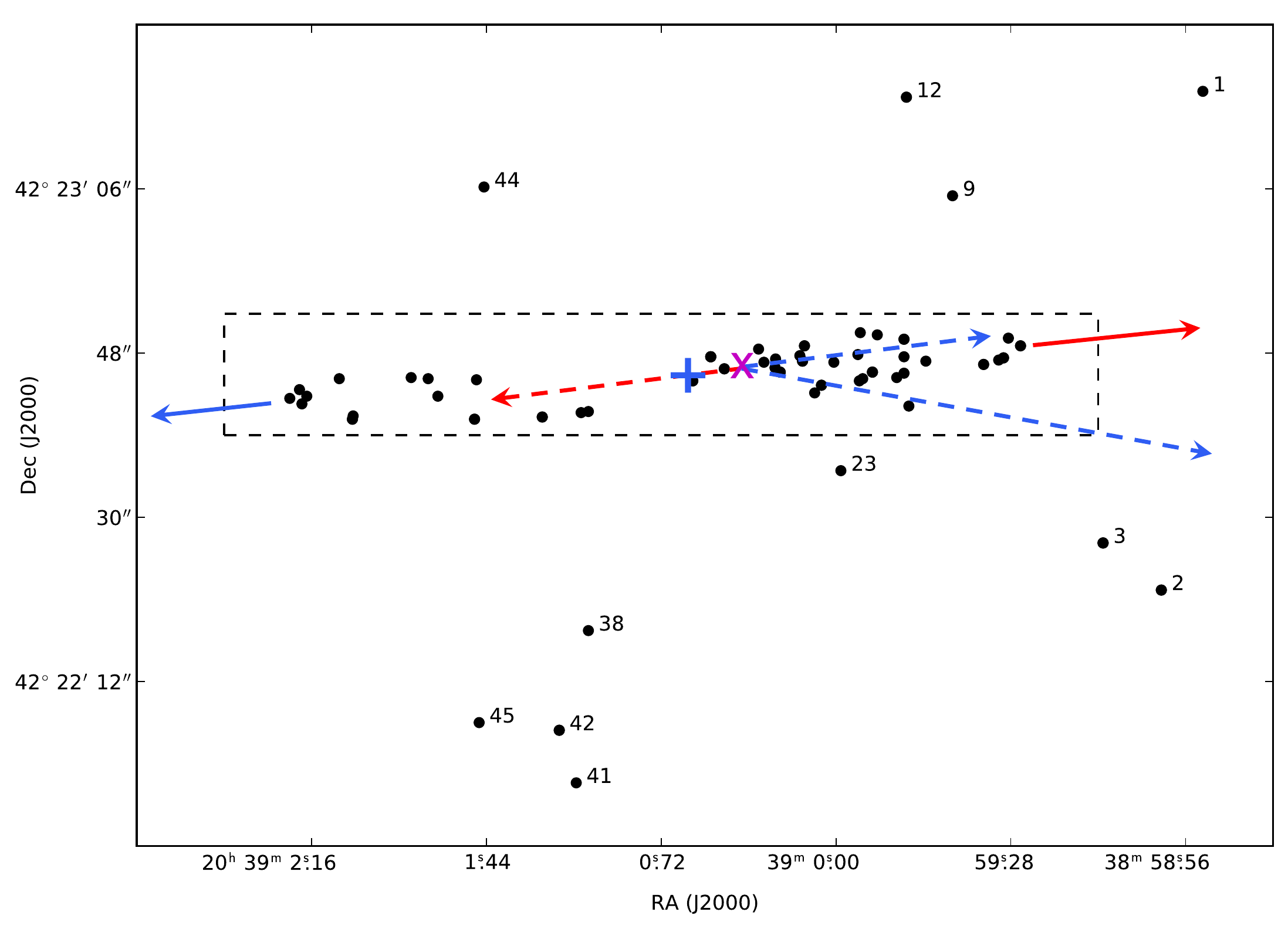} % FullMap2017.pdf}
\caption{The locations of the 44 GHz Class I \meth\ maser spots in the high mass star forming region \droh\ from the 2017 observations.
Of the 57 masers listed in Table~\ref{tab:2017masers}\ and shown in this figure, 46 are in the smaller region delineated by the dashed rectangle, and are shown in more detail in Figure~\ref{fig:2017smallmap}. The remaining 11 masers are to the north and south of the smaller region, and are numbered corresponding to their designation in Table~\ref{tab:2017masers}. The horizontal axis shows the Right Ascension (RA) and the vertical axis shows the Declination (Dec), both in epoch J2000. The blue plus sign shows the position of the 1.4 mm source SMA4 from \citet{zapata+2012}. The red and blue solid arrows mark the roughly east-west direction of the low-velocity 44~GHz \meth\ maser outflow; \citet{zapata+2012} proposed SMA4 as the source of this outflow. The purple cross marks the likely origin of the high-velocity CO outflows, which arise from a different source than the \meth\ maser outflow (Section~\ref{sec:intro}). Dashed blue and red arrows show one of the high-velocity CO ($J$ = 2$-$1) outflows, and the dashed blue arrow toward the southwest shows another monopolar CO ($J$ = 2$-$1) outflow.  \label{fig:2017map} }
\end{figure}

A total of 57 Class~I~\meth\ maser spots at 44 GHz were identified in the data from the 2017 observations of the high mass star forming region \droh. The location, peak intensity, center velocity, and Full Width at Half Maximum (FWHM) velocity linewidth of these masers are listed in Table~\ref{tab:2017masers}, and their distribution is shown in Figure~\ref{fig:2017map}. There are two masers with intensities higher than 100~\jybeam, 10 masers with intensities 10-100~\jybeam, 19 masers with intensities 1-10~\jybeam, and the remaining 26 have intensities lower than 1~\jybeam\ (Table~\ref{tab:2017masers}). The center velocities of these masers are found to range between $-$8.65 to +2.56 \kms.
FWHM linewidths range from 0.193 \kms\ to 0.892~\kms, with two exceptions; maser 37 has two components, one of which has a linewidth of 1.10~\kms, and maser 44 has a linewidth of 1.71~\kms. Of the 57 masers, 36 were fitted with single Gaussian components, and 21 required two or more Gaussian components.

\onecolumngrid

% Figure 1 (map of 2017 masers)

\startlongtable % --- needed for multi-page portrait table
\begin{deluxetable}{cccccc}
\tablecaption{Fitted parameters of the Observed 44 GHz Class I \meth\ Masers in \droh \label{tab:2017masers}}
\tablenum{2}
\tablehead{
\colhead{(1)} & \colhead{(2)} & \colhead{(3)} & \colhead{(4)}  & \colhead{(5)} & \colhead{(6)}  \\
	& \colhead{R.A.} & \colhead{Decl.} & \colhead{Intensity\tablenotemark{\rm \scriptsize a}}  & \colhead{Center Velocity} & \colhead{Velocity linewidth\tablenotemark{\rm \scriptsize b}} \\
		\colhead{Maser \#} & \colhead{(J2000)} & \colhead{(J2000)} & \colhead{(\jybeam)}  & \colhead{(\kms)} & \colhead{(\kms)}   }
\startdata
1	&	20	38	58.49	&	42	23	16.6	&	1.10	$\pm$	0.02	&	$-$2.207	$\pm$	0.003	&	0.312	$\pm$	0.008		\\
2	&	20	38	58.66	&	42	22	22.0	&	0.84	$\pm$	0.02	&	$-$1.266	$\pm$	0.003	&	0.350	$\pm$	0.008		\\
3a	&	20	38	58.89	&	42	22	27.2	&	0.45	$\pm$	0.01	&	$-$5.095	$\pm$	0.017	&	0.599	$\pm$	0.029	\\
3b	&	\nodata			&	\nodata				&	0.31	$\pm$	0.02	&	$-$5.802	$\pm$	0.025	&	0.769	$\pm$	0.084		\\
3c	&	\nodata			&	\nodata				&	0.17	$\pm$	0.01	&	$-$6.729	$\pm$	0.061	&	0.892	$\pm$	0.119		\\
4a	&	20	38	59.25	&	42	22	48.8	&	197.64 $\pm$	8.91	&	0.850	$\pm$	0.002	&	0.368	$\pm$	0.004		\\
4b	&	\nodata			&	\nodata				&	75.69 $\pm$	5.00	&	0.591	$\pm$	0.022	&	0.518	$\pm$	0.021		\\
5	&	20	38	59.29	&	42	22	49.6	&	13.65 $\pm$	0.13	&	1.140	$\pm$	0.001	&	0.312	$\pm$	0.004		\\
6a	&	20	38	59.31	&	42	22	47.5	&	6.60	$\pm$	0.24	&	$-$0.887	$\pm$	0.002	&	0.388	$\pm$	0.004		\\
6b	&	\nodata			&	\nodata				&	0.67	$\pm$	0.12	&	$-$1.188	$\pm$	0.081	&	0.596	$\pm$	0.084		\\
7	&	20	38	59.33	&	42	22	47.2	&	5.52	$\pm$	0.06	&	$-$1.019	$\pm$	0.002	&	0.344	$\pm$	0.005		\\
8	&	20	38	59.39	&	42	22	46.8	&	0.72	$\pm$	0.01	&	$-$1.416	$\pm$	0.002	&	0.350	$\pm$	0.006		\\
9	&	20	38	59.52	&	42	23	05.2	&	3.48	$\pm$	0.02	&	$-$0.833	$\pm$	0.001	&	0.241	$\pm$	0.001		\\
10	&	20	38	59.63	&	42	22	47.1	&	0.14	$\pm$	0.01	&	$-$0.347	$\pm$	0.011	&	0.533	$\pm$	0.028		\\
11	&	20	38	59.70	&	42	22	42.2	&	0.27	$\pm$	0.01	&	$-$0.112	$\pm$	0.050	&	0.486	$\pm$	0.015		\\
12a	&	20	38	59.71	&	42	23	16.0	&	0.73	$\pm$	0.01	&	$-$0.926	$\pm$	0.008	&	0.585	$\pm$	0.019		\\
12b	&	\nodata			&	\nodata				&	0.51	$\pm$	0.02	&	$-$0.350	$\pm$	0.009	&	0.386	$\pm$	0.018	\\
13	&	20	38	59.72	&	42	22	45.8	&	0.45	$\pm$	0.01	&	$-$0.078	$\pm$	0.004	&	0.541	$\pm$	0.011		\\
14	&	20	38	59.72	&	42	22	47.6	&	0.31	$\pm$	0.01	&	0.142	$\pm$	0.004	&	0.193	$\pm$	0.009		\\
15a	&	20	38	59.72	&	42	22	49.5	&	1.01	$\pm$	0.05	&	0.216	$\pm$	0.003	&	0.225	$\pm$	0.014		\\
15b	&	\nodata			&	\nodata				&	0.45	$\pm$	0.04	&	0.015	$\pm$	0.016	&	0.408	$\pm$	0.017		\\
16	&	20	38	59.75	&	42	22	45.3	&	0.57	$\pm$	0.01	&	$-$0.265	$\pm$	0.004	&	0.362	$\pm$	0.010		\\
17	&	20	38	59.83	&	42	22	50.0	&	0.19	$\pm$	0.01	&	$-$1.038	$\pm$	0.010	&	0.555	$\pm$	0.027		\\
18a	&	20	38	59.85	&	42	22	45.9	&	12.01$\pm$	0.43	&	$-$0.214	$\pm$	0.001	&	0.205	$\pm$	0.003		\\
18b	&	\nodata			&	\nodata				&	2.78	$\pm$	0.22	&	$-$0.464	$\pm$	0.007	&	0.263	$\pm$	0.015		\\
18c	&	\nodata			&	\nodata				&	2.40	$\pm$	0.25	&	$-$0.030	$\pm$	0.051	&	0.470	$\pm$	0.059		\\
19	&	20	38	59.89	&	42	22	45.2	&	18.42 $\pm$	0.44	&	0.337	$\pm$	0.004	&	0.226	$\pm$	0.005		\\
20	&	20	38	59.90	&	42	22	44.9	&	16.27 $\pm$	0.33	&	0.561	$\pm$	0.002	&	0.203	$\pm$	0.002		\\
21	&	20	38	59.90	&	42	22	50.2	&	0.19	$\pm$	0.01	&	$-$0.799	$\pm$	0.010	&	0.489	$\pm$	0.028		\\
22	&	20	38	59.91	&	42	22	47.8	&	0.16	$\pm$	0.01	&	$-$0.669	$\pm$	0.010	&	0.598	$\pm$	0.028		\\
23	&	20	38	59.97	&	42	22	35.1	&	0.60	$\pm$	0.01	&	$-$1.830	$\pm$	0.002	&	0.250	$\pm$	0.005		\\
24a	&	20	39	0.02	&	42	22	47.0		&	0.26 $\pm$	0.01	&	$-$1.673 	$\pm$	0.016	&	0.481	$\pm$	0.027		\\
24b	&	\nodata	&	\nodata		&	0.11 $\pm$	0.01	&	$-$1.103 	$\pm$	0.042	&	0.569	$\pm$	0.082		\\
25a	&	20	39	0.06	&	42	22	44.5		&	1.67	$\pm$	0.03	&	$-$8.650	$\pm$	0.001	&	0.326	$\pm$	0.004		\\
25b	&	\nodata			&	\nodata				&	0.58	$\pm$	0.03	&	$-$8.490	$\pm$	0.010	&	0.630	$\pm$	0.011		\\
26	&	20	39	0.09	&	42	22	43.6		&	1.43	$\pm$	0.01	&	$-$5.112	$\pm$	0.002	&	0.353	$\pm$	0.004		\\
27a	&	20	39	0.13	&	42	22	48.8		&	0.09	$\pm$	0.02	&	$-$2.463	$\pm$	0.158	&	0.812	$\pm$	0.283		\\
27b	&	\nodata			&	\nodata				&	0.11	$\pm$	0.05	&	$-$1.973	$\pm$	0.037	&	0.487	$\pm$	0.082		\\
28a	&	20	39	0.14	&	42	22	47.1		&	1.07	$\pm$	0.03	&	$-$1.321	$\pm$	0.006	&	0.562	$\pm$	0.009		\\
28b &	\nodata			&	\nodata		&	0.41	$\pm$	0.01	&	$-$0.720	$\pm$	0.026	&	0.791	$\pm$	0.042		\\
29	&	20	39	0.15	&	42	22	47.7		&	10.07 $\pm$	0.10	&	$-$1.445	$\pm$	0.002	&	0.395	$\pm$	0.005	\\
30a	&	20	39	0.23	&	42	22	45.9		&	277.66 $\pm$	3.89	&	0.465	$\pm$	0.002	&	0.279	$\pm$	0.003		\\
30b	&	\nodata			&	\nodata				&	57.42 $\pm$	2.75	&	0.205	$\pm$	0.013	&	0.316	$\pm$	0.018		\\
31a	&	20	39	0.25	&	42	22	46.4		&	3.59	$\pm$	0.12	&	0.323	$\pm$	0.007	&	0.472	$\pm$	0.006		\\
31b  &	\nodata			&	\nodata				&	0.83	$\pm$	0.02	&	1.182	$\pm$	0.008	&	0.570	$\pm$	0.021		\\
32	&	20	39	0.25	&	42	22	47.4		&	12.40	$\pm$	0.11	&	0.114	$\pm$	0.004	&	0.511	$\pm$	0.008	\\
33a	&	20	39	0.30	&	42	22	47.0		&	1.23	$\pm$	0.01	&	$-$2.099	$\pm$	0.002	&	0.366	$\pm$	0.005	\\
33b	&	\nodata			&	\nodata				&	0.28	$\pm$	0.01	&	$-$2.786	$\pm$	0.009	&	0.417	$\pm$	0.023		\\
34	&	20	39	0.32	&	42	22	48.4		&	1.10	$\pm$	0.02	&	$-$0.046	$\pm$	0.005	&	0.451	$\pm$	0.012		\\
35a	&	20	39	0.46	&	42	22	46.3		&	0.10	$\pm$	0.01	&	$-$2.482	$\pm$	0.014	&	0.619	$\pm$	0.037		\\
35b	&	\nodata			&	\nodata				&	0.09	$\pm$	0.01	&	$-$0.699	$\pm$	0.011	&	0.342	$\pm$	0.031		\\
36a	&	20	39	0.52	&	42	22	47.6		&	1.60	$\pm$	0.05	&	2.561	$\pm$	0.010	&	0.398	$\pm$	0.015		\\
36b	&	\nodata			&	\nodata				&	1.23	$\pm$	0.06	&	2.137	$\pm$	0.013	&	0.396	$\pm$	0.051		\\
36c &	\nodata			&	\nodata				&	0.55	$\pm$	0.02	&	1.357	$\pm$	0.008	&	0.289	$\pm$	0.016		\\
36d	&	\nodata			&	\nodata				&	2.37	$\pm$	0.08	&	1.774	$\pm$	0.007	&	0.344	$\pm$	0.013		\\
37a	&	20	39	0.59	&	42	22	44.9		&	0.09	$\pm$	0.01	&	$-$2.384	$\pm$	0.022	&	0.538	$\pm$	0.065		\\
37b	&	\nodata			&	\nodata				&	0.10	$\pm$	0.01	&	$-$1.577	$\pm$	0.039	&	1.097	$\pm$	0.119		\\
38	&	20	39	1.02	&	42	22	17.6		&	0.38	$\pm$	0.01	&	$-$3.182	$\pm$	0.007	&	0.377	$\pm$	0.017		\\
39	&	20	39	1.02	&	42	22	41.6		&	0.66	$\pm$	0.01	&	$-$1.049	$\pm$	0.002	&	0.469	$\pm$	0.006		\\
40	&	20	39	1.05	&	42	22	41.5		&	0.28	$\pm$	0.01	&	$-$1.175	$\pm$	0.007	&	0.491	$\pm$	0.017		\\
41	&	20	39	1.07	&	42	22	00.9 		&	1.56	$\pm$	0.05	&	$-$1.811	$\pm$	0.005	&	0.358	$\pm$	0.013		\\
42a	&	20	39	1.14	&	42	22	06.7		&	1.07	$\pm$	0.02	&	$-$1.034	$\pm$	0.005	&	0.750	$\pm$	0.016		\\
42b	&	\nodata			&	\nodata				&	0.54	$\pm$	0.03	&	$-$3.549	$\pm$	0.007	&	0.276	$\pm$	0.016		\\
43a	&	20	39	1.21	&	42	22	41.0		&	0.56	$\pm$	0.01	&	$-$2.349	$\pm$	0.011	&	0.711	$\pm$	0.018		\\
43b	&	\nodata			&	\nodata 				&	0.21	$\pm$	0.01	&	$-$2.933	$\pm$	0.019	&	0.519	$\pm$	0.030		\\
44	&	20	39	1.44	&	42	23	06.1		&	0.12	$\pm$	0.01	&	$-$2.020	$\pm$	0.029	&	1.709	$\pm$	0.101		\\
45	&	20	39	1.47	&	42	22	07.5   	&	0.16	$\pm$	0.02	&	$-$4.177	$\pm$	0.040	&	0.680	$\pm$	0.098		\\
46	&	20	39	1.48	&	42	22	45.1		&	1.22	$\pm$	0.01	&	$-$2.832	$\pm$	0.003	&	0.613	$\pm$	0.009		\\
47a	&	20	39	1.49	&	42	22	40.8		&	10.20 $\pm$	0.09	&	$-$4.380	$\pm$	0.002	&	0.387	$\pm$	0.006		\\
47b	&	\nodata			&	\nodata				&	1.85	$\pm$	0.46	&	$-$4.795	$\pm$	0.010	&	0.196	$\pm$	0.033		\\
48	&	20	39	1.64	&	42	22	43.3		&	12.25 $\pm$	0.04	&	$-$6.183	$\pm$	0.001	&	0.227	$\pm$	0.001		\\
49	&	20	39	1.67	&	42	22	45.2		&	1.31	$\pm$	0.01	&	$-$4.340	$\pm$	0.001	&	0.228	$\pm$	0.003		\\
50	&	20	39	1.75	&	42	22	45.3		&	0.14	$\pm$	0.01	&	$-$3.165	$\pm$	0.008	&	0.627	$\pm$	0.033		\\
51a	&	20	39	1.99	&	42	22	40.8		&	5.22	$\pm$	0.06	&	$-$5.210	$\pm$	0.005	&	0.428	$\pm$	0.007		\\
51b	&	\nodata			&	\nodata				&	3.34	$\pm$	0.09	&	$-$5.570	$\pm$	0.006	&	0.378	$\pm$	0.007		\\
52	&	20	39	1.99	&	42	22	41.1		&	5.07	$\pm$	0.02	&	$-$4.176	$\pm$	0.001	&	0.460	$\pm$	0.003		\\
53	&	20	39	2.05	&	42	22	45.2		&	16.56 $\pm$	0.97	&	$-$3.183	$\pm$	0.001	&	0.425	$\pm$	0.003		\\
54	&	20	39	2.18	&	42	22	43.3		&	0.65	$\pm$	0.02	&	$-$3.851	$\pm$	0.006	&	0.403	$\pm$	0.014		\\
55	&	20	39	2.20	&	42	22	42.4		&	4.09	$\pm$	0.05	&	$-$3.254	$\pm$	0.002	&	0.381	$\pm$	0.006		\\
56a	&	20	39	2.21	&	42	22	44.0		&	15.03 $\pm$	0.03	&	$-$5.039	$\pm$	0.001	&	0.617	$\pm$	0.002		\\
56b	&	\nodata			&	\nodata				&	2.74	$\pm$	0.03	&	$-$4.326	$\pm$	0.006	&	0.628	$\pm$	0.014		\\
57	&	20	39	2.25	&	42	22	43.0		&	0.56	$\pm$	0.01	&	$-$4.097	$\pm$	0.006	&	0.577	$\pm$	0.014		\\
\enddata
\tablenotetext{\rm a}{~The intensity values are primary beam corrected and are reported at the peak of the spectral line.}
\tablenotetext{\rm b}{~The velocity linewidth was measured at FWHM.}
\end{deluxetable}

\twocolumngrid

\begin{figure}[htb!]
\epsscale{0.85}
\plotone{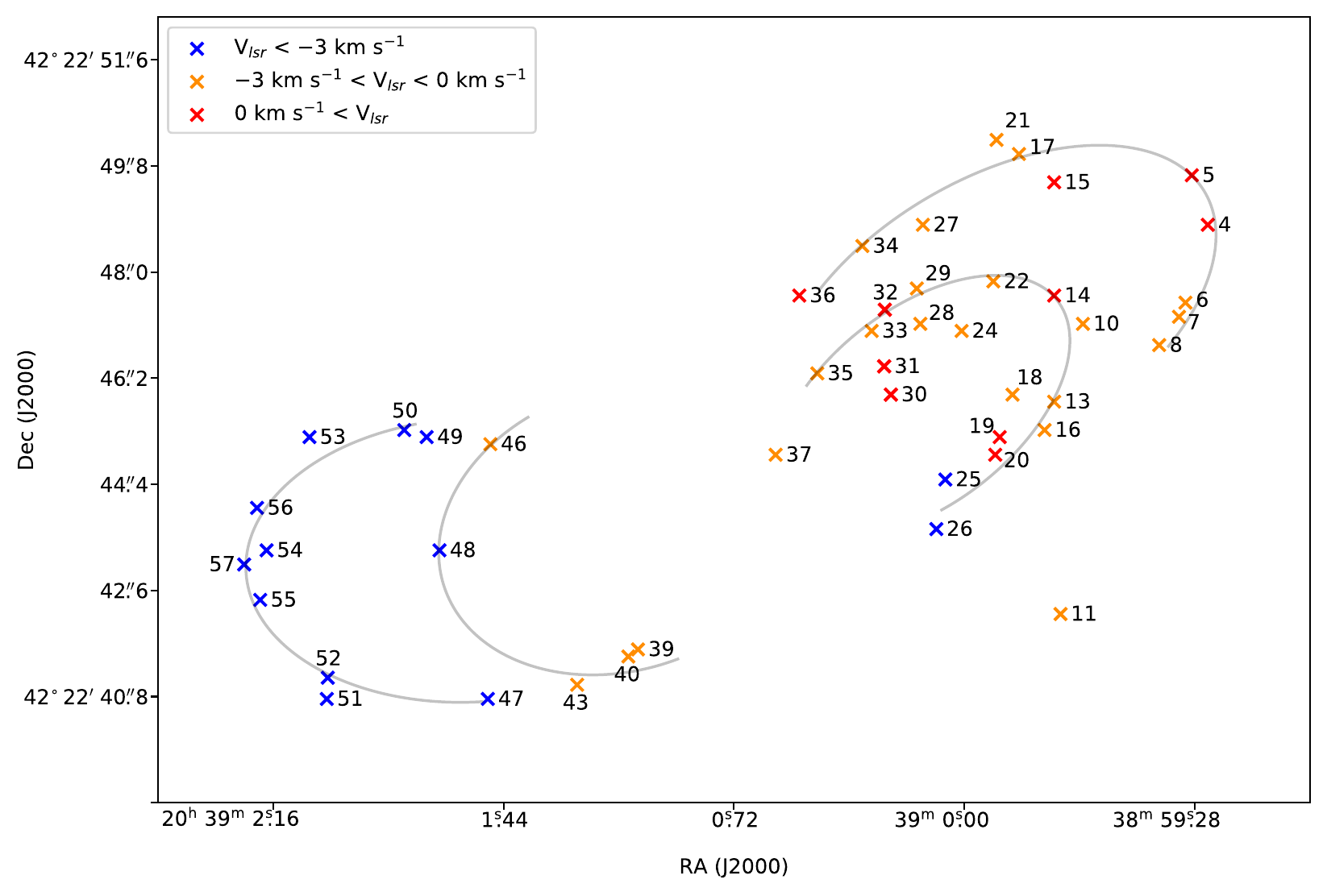}%NewSmallMap2017.pdf}
\caption{The masers in the region enclosed by the dashed rectangle in Figure~\ref{fig:2017map}. The numbers shown beside the position of each maser correspond to their designation in Table~\ref{tab:2017masers}. The red crosses show redshifted masers with $v_{\text{LSR}} > 0$~\kms, orange crosses show redshifted masers with $-3$~\kms~$< v_{\text{LSR}} < 0$~\kms, and blue crosses show blueshifted masers with $v_{\text{LSR}} < -3$~\kms. The systemic velocity of \droh\ is $-$3~\kms\ (\citealt{chandler+1993}). To guide the eye, thin lines have been used to mark the outer and inner arcs in the western and eastern lobes.  \label{fig:2017smallmap} }
\end{figure}

\begin{figure}[htb!]
\epsscale{0.85}
\plotone{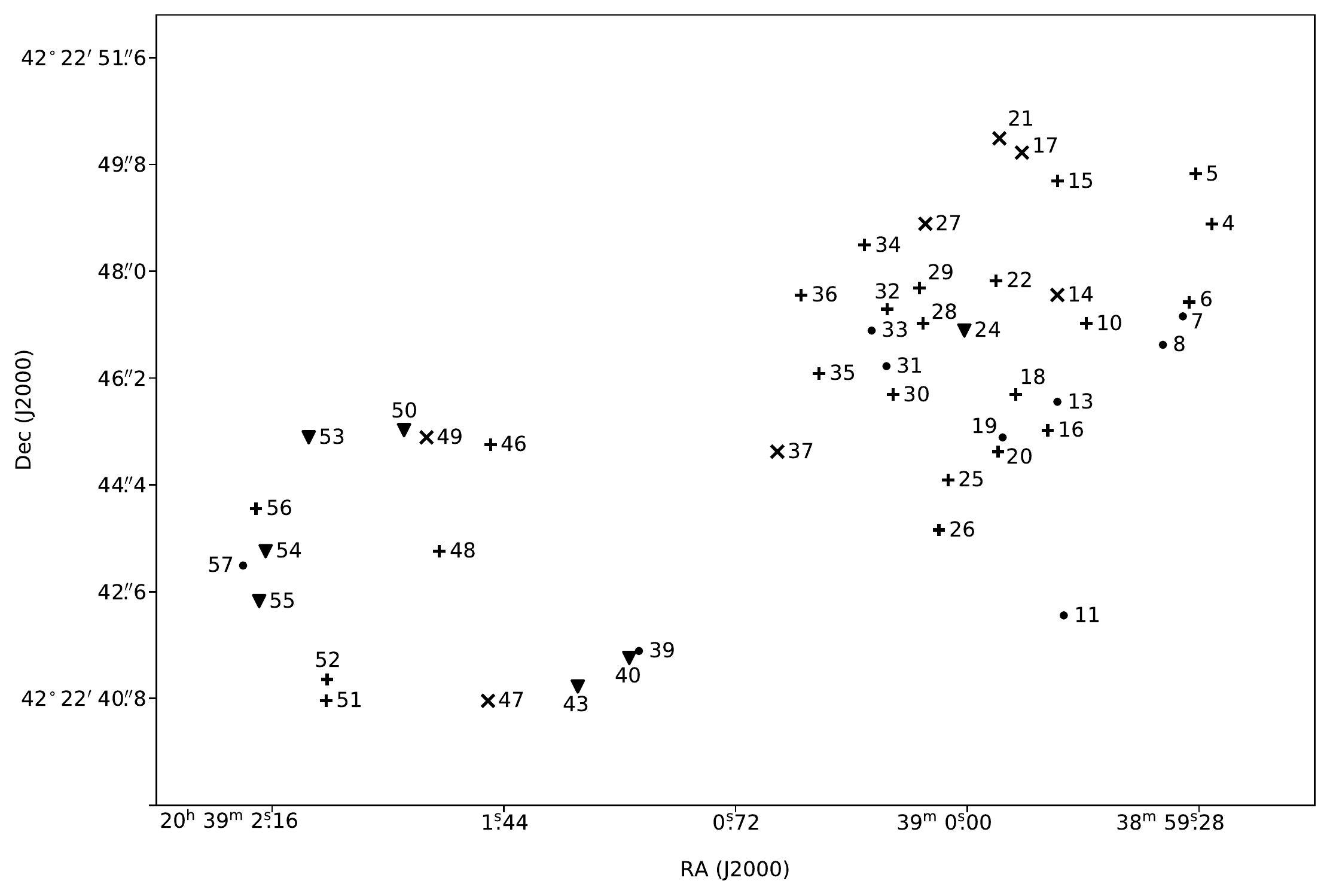}
\caption{The same image as in Figure~\ref{fig:2017smallmap}, but now showing which masers have increased or decreased in intensity from 2001 to 2017. Plus signs indicate masers whose intensity has increased in our low resolution comparison of 2001 and 2017 data, upside down triangles mark masers whose intensity has decreased from 2001 to 2017, crosses indicate masers for which there was no significant change in intensity (above our adopted threshold of 10\%), and dots indicate masers observed in 2017 that were not observed in 2001. \label{fig:2017changesMap} }
\end{figure}

Of the 57 maser spots found in \droh, 46 are located in a smaller region of size\footnote{At the adopted distance of 1.5~kpc \citep{rygl+2012} to \droh, 1\arcsec\ $\equiv$ 0.00727~pc.} 0.3~pc $\times$ 0.1~pc (enclosed by the dashed rectangle in Figure~\ref{fig:2017map}), whereas the remaining 11 masers are located to the north and south of this region. The region enclosed by the dashed rectangle is shown in Figure~\ref{fig:2017smallmap}. In agreement with \citet{kurtz+2004}, the masers are seen to be distributed in a western and an eastern lobe in this figure, with 31 masers located in the western lobe and 15 in the eastern lobe. Figure~\ref{fig:2017smallmap} also reveals that the masers in both the western and eastern lobes are arranged in two arc-like structures in each lobe, in agreement with \citet{araya+2009}. The list of masers in each arc is given in Table~\ref{tab:ArcMasers}, and the arcs are marked in Figure~\ref{fig:2017smallmap}. Figure~\ref{fig:2017changesMap} shows the same region as in Figure~\ref{fig:2017smallmap}, except that the masers are marked with different symbols to depict intensity changes from 2001 to 2017. The outer arc in the western lobe is comprised of 11 masers, whereas there are 20 masers in the inner arc. There are fewer masers in the eastern lobe, with only 5 masers in the inner arc and 10 in the outer arc.  Note that we have chosen to put masers 27, 34, and 36 (Table~\ref{tab:ArcMasers} and Figure~\ref{fig:2017smallmap}) in the outer arc of the western lobe, unlike \citet{araya+2009} who put them in the inner arc; this gives a better representation of bow shocks driven by collimated outflows (see, e.g., \citealt{arce+2007}). Irrespective of this choice (to put masers 27, 34, and 36 in the inner or outer arc of the western lobe), the observed velocity profile is consistent with an arc-like structure, with high velocities at the head of the bow shock, in both arcs of the western and eastern lobes. The shift in the position angle of the eastern lobe seen in Figure~\ref{fig:2017smallmap} was also observed by \citet{araya+2009}, who speculated that it might arise from jet precession combined with inhomogeneities in the molecular gas interacting with the outflow. There are almost as many masers fitted with single components as there are with multiple Gaussian components in both the inner and outer arcs of the western lobe (with 6  single component masers out of 11 in the outer arc, and 11 out of 20 in the inner arc). In the eastern lobe, however, significantly more masers were fitted with single components than with multiple Gaussian components (with 4 single component masers out of 5 in the inner arc, and 7 out of 10 in the outer arc). There is no discernible pattern in the locations along the arcs where masers with single components are found compared to masers with multiple Gaussian components. For example, masers 4 and 6 in the outer arc of the western lobe each have two Gaussian components and masers 5, 7, and 8 each have a single component; maser 5 is located near maser 4, and masers 7 and 8 are near maser 6. Compared to the $-$3~\kms\ systemic velocity of \droh\ reported in \citet{chandler+1993}, almost all the masers in the western lobe are redshifted. Their center velocities range from 2.561~\kms\ to $-$2.786~\kms, with the exception of two masers; masers 25 and 26 are blueshifted. In the eastern lobe, the masers are largely blueshifted. Their center velocities range from $-$3.165~\kms\ to $-6.183$~\kms, with the exception of masers 39, 40, 43, and 46, which are redshifted. Of the 46 masers located in the western and eastern lobes, 33 are redshifted and 13 are blueshifted. Of the 33 redshifted masers, 19 were fitted with single components, and 14 with multiple Gaussian components. Of the 13 blueshifted masers, 9 were fitted with single components and 4 with multiple Gaussian components.

\onecolumngrid

% Figure 2 (smaller extent map of 2017 masers)

%\vsp{0.1}

%\onecolumngrid

% Figure 3 (increase/decrease/unchanged masers from 2001 to 2017 --- new, based on referee comment)

%\twocolumngrid

\twocolumngrid

\clearpage

\begin{figure}[htb!]
\epsscale{0.9}
\plotone{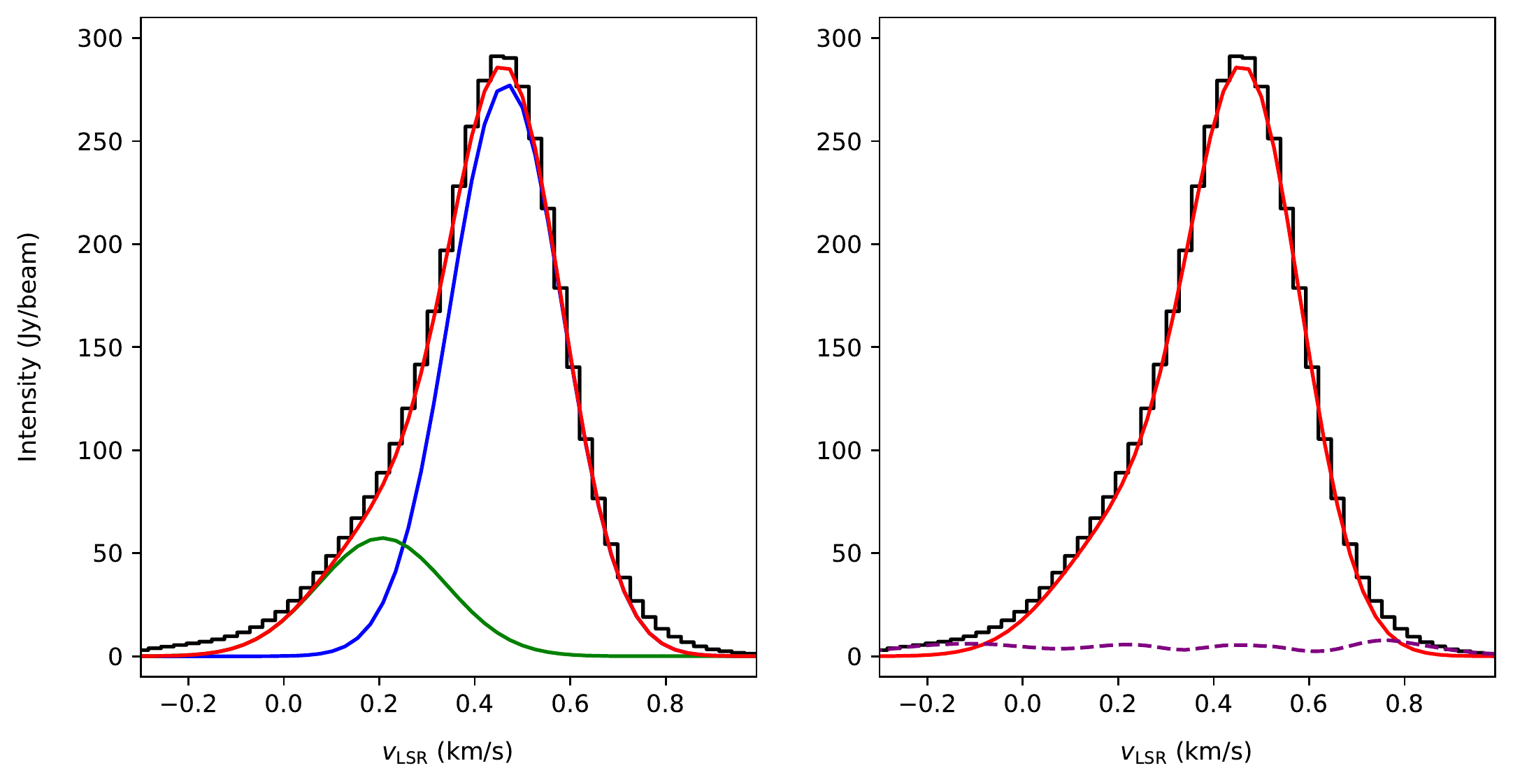}
\caption{Observed spectral profile (black histogram--like line) at 44 GHz of the Class~I \meth\ line toward DR21(OH) designated as maser 30 in Table~\ref{tab:2017masers}. Maser~30 is the strongest maser in \droh, and is located in the inner arc of the western lobe (Figure~\ref{fig:2017smallmap}). The blue and green curves in the left panel illustrate the Gaussian components fitted to this observed profile. The red curve (in both panels) is the sum of the two Gaussian components shown in the left panel. The dashed purple curve in the right panel is the residual from the fit. Note that since both panels have the same $y$-axis, the $y$-axis in the right panel has not been labeled to avoid clutter.  \label{fig:mas30} }
\end{figure}

\begin{figure}[htb!]
\epsscale{1.0}
\plotone{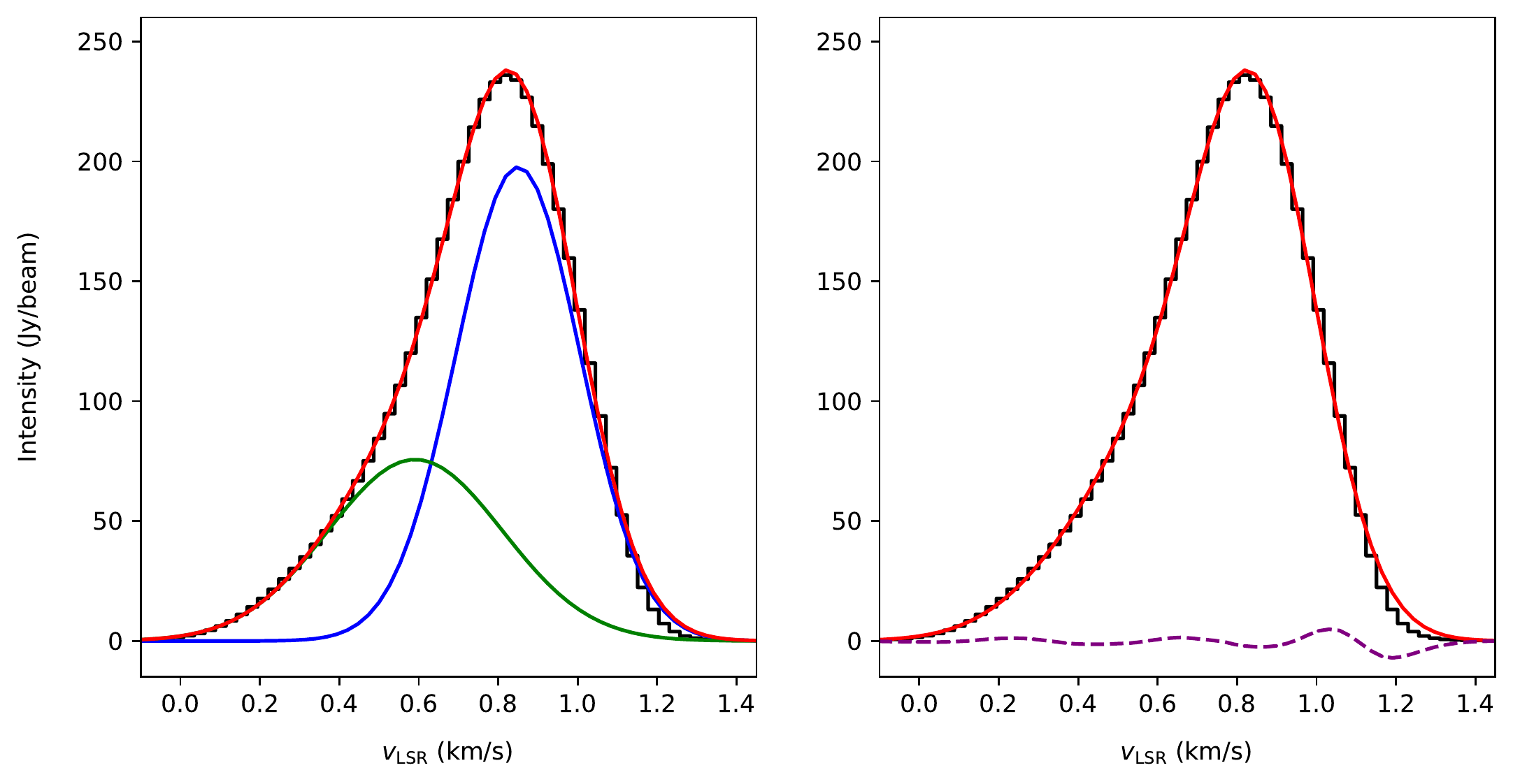}
\caption{Observed spectral profile (black histogram--like line) at 44 GHz of the Class~I \meth\ line toward DR21(OH) designated as maser 4 in Table~\ref{tab:2017masers}. Maser~4 is the second strongest maser in \droh, and is located in the outer arc of the western lobe (Figure~\ref{fig:2017smallmap}). The blue and green curves in the left panel illustrate the Gaussian components fitted to this observed profile. The red curve (in both panels) is the sum of the two Gaussian components shown in the left panel. The dashed purple curve in the right panel is the residual from the fit. \label{fig:mas4} }
\end{figure}

\begin{figure}[htb!]
\epsscale{1.0}
\plottwo{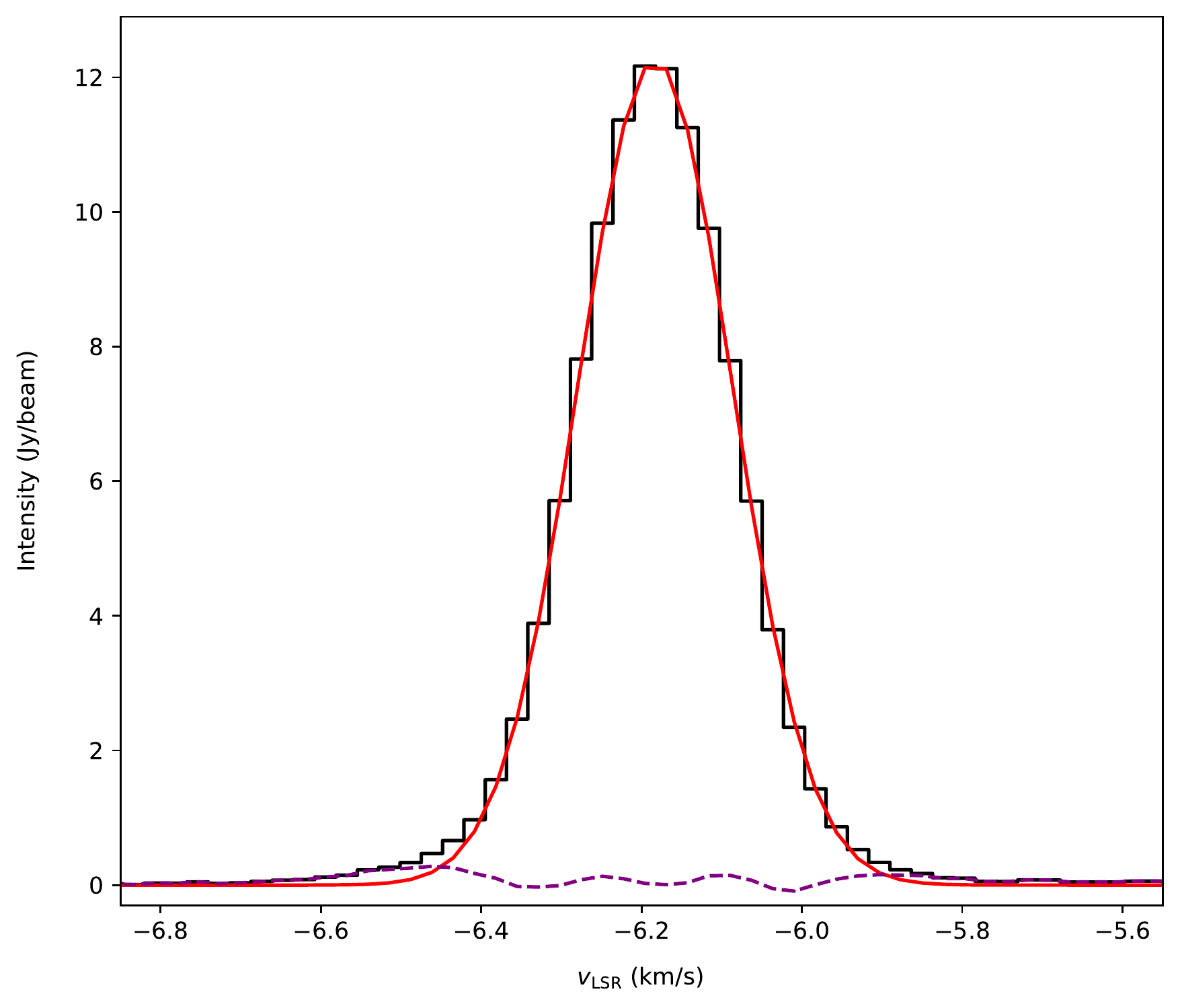}{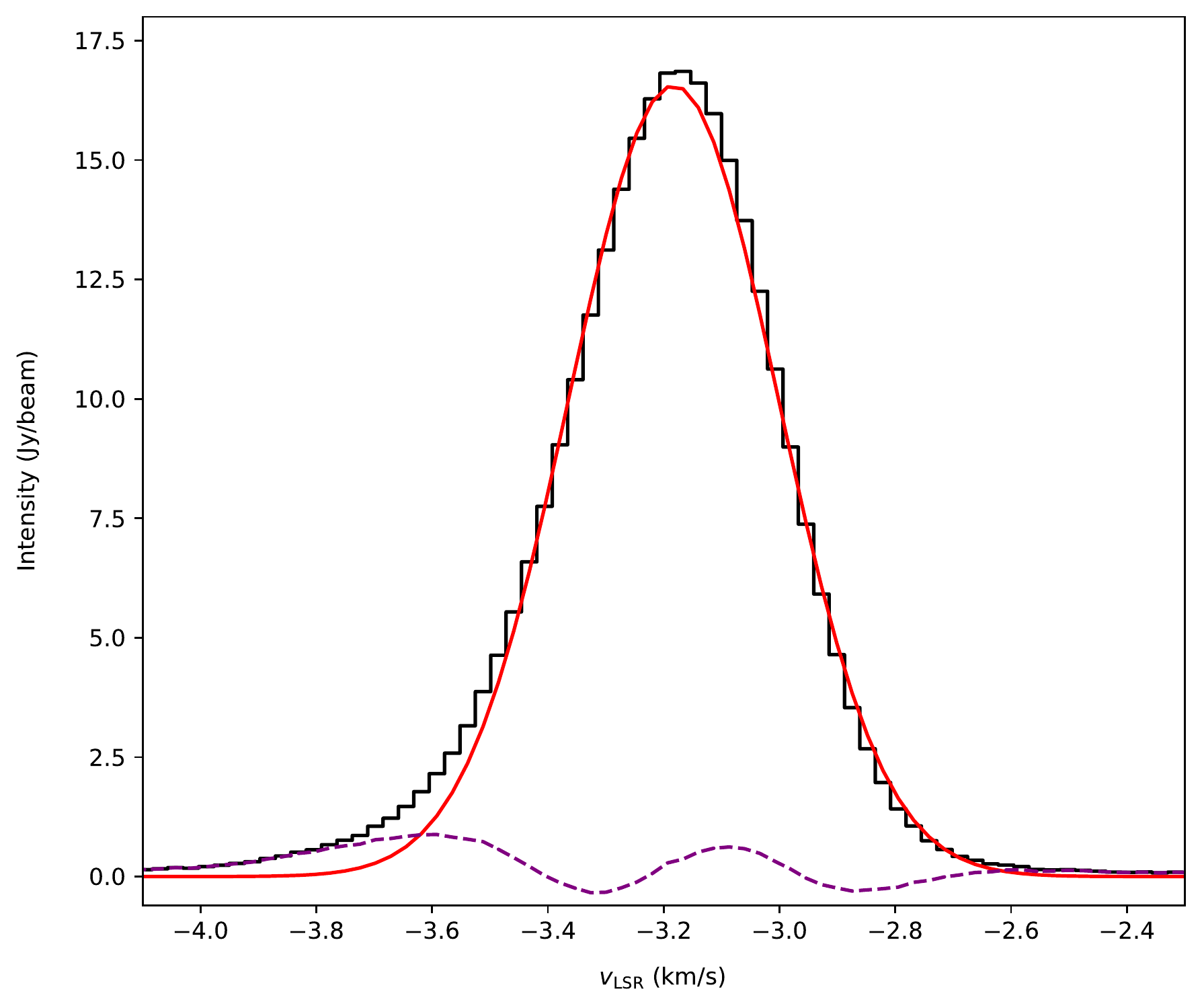}

\caption{Observed spectral profiles (black histogram--like line) at 44 GHz of the Class~I \meth\ line toward DR21(OH) designated as maser 48 (on the left) and maser 53 (on the right) in Table~\ref{tab:2017masers}. Maser~48 is the strongest maser in the inner arc of the eastern lobe, and maser~53 is the strongest maser in the outer arc of the eastern lobe (Figure~\ref{fig:2017smallmap}). Both profiles were fitted by a single Gaussian component shown by the red curve; the dashed purple curve shows the residuals from the fit.\label{fig:mas48} }
\end{figure}

The strongest maser in DR21(OH), maser 30 in Table~\ref{tab:2017masers}, is located in the inner arc of the western lobe (Figure~\ref{fig:2017smallmap} and Table~\ref{tab:ArcMasers}). The observed profile for maser 30 is shown in Figure~\ref{fig:mas30}, together with the two fitted Gaussian components, fitted profile, and residual from the fit. The stronger and narrower component (FWHM velocity linewidth 0.279 \kms) has an intensity of 278~\jybeam\ and is centered at 0.465 \kms, whereas the lower intensity (57 \jybeam) and slightly broader component (FWHM velocity linewidth 0.316 \kms) is centered at 0.205 \kms. The second strongest maser, maser 4 in Table~\ref{tab:2017masers}, is located in the outer arc of the western lobe. The fits to the observed profile for maser 4 shown in Figure~\ref{fig:mas4} also reveal two masers blended in velocity; the stronger (198 \jybeam) and narrower component (FWHM velocity linewidth 0.368 \kms) is centered at 0.850 \kms, whereas the lower intensity (76 \jybeam) and slightly broader component (FWHM velocity linewidth 0.518 \kms) is centered at 0.591 \kms. Also shown are the strongest maser in the inner arc of the eastern lobe (maser 48), and maser 53, the strongest maser in the outer arc of the eastern lobe (Figure~\ref{fig:mas48}). Maser 48 was fitted with a single narrow Gaussian component (FWHM velocity linewidth 0.227~\kms) of intensity 12.3~\jybeam\ centered at $-$6.183~\kms. Maser 53 was also fitted with a single Gaussian component of intensity 16.6~\jybeam\ and FWHM linewidth 0.425~\kms, centered at $-$3.183~\kms.

\onecolumngrid

% Figure 4 (maser 30)

% Figure 5 (maser 4)

\begin{deluxetable}{lcccl}
\tablenum{3}
\tablecaption{Masers in the inner and outer arcs of the western and eastern lobes \label{tab:ArcMasers}} 
\tablehead{
	\colhead{} && \colhead{Number} && \colhead{List of} \\
	\colhead{Location} && \colhead{of masers} && \colhead{Masers} }
\startdata
Outer arc, western lobe && 11 && 4, 5, 6, 7, 8, 15, 17, 21, 27, 34, 36 \\ \hline
Inner arc, western lobe && 20 && 10, 11, 13, 14, 16, 18, 19, 20, 22, 24, \\
	&&  && 25, 26, 28, 29, 30, 31, 32, 33, 35, 37 \\ \hline
Inner arc, eastern lobe && 5 && 39, 40, 43, 46, 48 \\ \hline
Outer arc, eastern lobe && 10 & & 47, 49, 50, 51, 52, 53, 54, 55, 56, 57 \\
\enddata
\end{deluxetable}

%\clearpage

% Figure 6 (maser 48) --- combined with maser 53 in revised version

\twocolumngrid

\clearpage

\clearpage

\onecolumngrid

\begin{deluxetable}{lccccccccccl}
\tablenum{4}
\tablewidth{0pt}
\tablecaption{Observations discussed in this paper
	\protect\label{tCOBS}}
\tablehead{
\colhead{Parameter} &  \multicolumn{3}{c}{Values} }
\startdata
Date of observation \dotfill & 2001 Sep 7\tablenotemark{a} & 2012 Apr 25\tablenotemark{b} & 2017 May 24    \\
Configuration \dotfill & C & C & C \\
Angular Resolution \dotfill & $\sim 0\, \rlap{\arcsec}.\, 6$ &  $\sim 0\, \rlap{\arcsec}.\, 6$ &  $\sim 0\, \rlap{\arcsec}.\, 6$ \\ 
Velocity Resolution (\kms) \ldots\dotfill & 0.66 & 0.03 & 0.03 \\
Total bandwidth (\kms) \dotfill & 21\tablenotemark{c} & 6.8\tablenotemark{c} & 27\tablenotemark{c} \\
\enddata
%\tablecomments{}
\tablenotetext{a}{ From \citet{araya+2009}.}
\tablenotetext{b}{ From data published in \citet{momjian+2017}.}
\tablenotetext{c}{ The 2001 and 2017 bandwidths are more than adequate to observe all the masers in \droh, whose velocities range from $-$8.65 to +2.56~\kms, a velocity range of 11.2~\kms. The 2012 observations targeted a subset of these masers in the velocity range from $-$2.369 to 2.524~\kms.}
\end{deluxetable}

\twocolumngrid

\section{Discussion} \label{sec:disc}

The maser spots in \droh\ appear to be stable entities, with the overwhelming majority of them being present from 2001 through 2017, but the intensities of several do appear to change. We begin with a discussion of the maser distribution across all three epochs in \S~\ref{sec:masDis}. The variability of masers is then discussed in \S~\ref{sec:var}, and a discussion on the likely causes of these variations is presented in \S~\ref{sec:reasons}. We have limited our discussion in this paper to variations in the intensity of the masers.

\vsp{0.05}

\subsection{Maser Distribution} \label{sec:masDis}

We detected 57 maser spots in the 2017 observations. In their 2001 observations, \citet{araya+2009} reported a total of 49 masers. Their observations were also carried out with the VLA in the C configuration and have a similar angular resolution (0.6\arcsec) as the 2017 observations; Table~\ref{tCOBS} lists the relevant parameters for the 2001, 2012, and 2017 observations. The positions of the masers can therefore be directly compared between these two epochs. Only one of the masers from 2001 reported by \citet{araya+2009} was not detected in the 2017 observations. The positions of the other 48 masers  are listed in Table~\ref{tab:2001masers}. Also listed in this table is the corresponding number of the maser in the 2017 observations, taken from Table \ref{tab:2017masers}. Such correspondence was established by direct comparison of the positions (in RA and Dec) of the masers in the 2001 and 2017 observations, while allowing for differences in position within the beam width (0.6\arcsec). The spectral resolution of the 2001 data, however, is much lower (0.66~\kms\ compared to 0.0266~\kms\ in 2017). Thus, of the 48 masers in Table~\ref{tab:2001masers}, maser 2001-5 could be masers 6 and 7 in the 2017 observations (Table~\ref{tab:2017masers}). \citet{araya+2009} might have observed these two as only one maser due to their lower velocity resolution. Likewise, maser 2001-16 in Table~\ref{tab:2001masers} could be masers 19 and 20 in the 2017 observations, which might again have showed up as one maser due to the lower velocity resolution in the 2001 data. The remaining seven masers in the 2017 data are low intensity masers that were not reported by \citet{araya+2009}. In principle, though, it is possible that all nine of these could be new masers that did not exist in 2001. Nevertheless, it is clear that Class~I \meth\ masers are stable entities, since almost 90\% of the masers have appeared at the same position from 2001 to 2017.

\vsp{0.1}

A subset of these masers was observed in 2012 with the aim of measuring the Zeeman effect \citep{momjian+2017}. These observations too were carried out with the VLA in the C configuration, and have a similar angular resolution and the same spectral resolution as the 2017 observations reported in this paper. Therefore, the 2012 observations can be directly compared to the 2017 observations. However, the frequency coverage of the 2012 observations was smaller (1~MHz compared to 4~MHz in 2017), and thus only 24 maser spots were observed. The positions of these masers, along with other pertinent information taken from \citet{momjian+2017}, are listed in Table~\ref{tab:2012masers}. Also listed in this table are the corresponding numbers of the masers in the 2017 observations from Table~\ref{tab:2017masers}. All 24 of the masers observed in 2012 were also detected in the 2017 observations. The center velocities of the masers observed in 2012 are in the range $-$2.369 \kms\ to 2.524 \kms. 

%\onecolumngrid

\startlongtable
\begin{deluxetable}{cccc}
\tablenum{5}
\tablecaption{Observed Masers in \droh\ from \citet{araya+2009} \label{tab:2001masers}} 
\tablehead{\colhead{(1)} &  \colhead{(2)} &  \colhead{(3)} & \colhead{(4)}  \\
 	\colhead{Araya} & \colhead{R.A.} & \colhead{Decl.} & \colhead{Corresponding} \\
	\colhead{Maser ID} & \colhead{(J2000)} & \colhead{(J2000)} & \colhead{2017 maser} }
\startdata
2001$-$1	&	20	38	58.68	&	42	22	22.1	&	2	\\
2001$-$2	&	20	38	58.92	&	42	22	27.2	&	3	\\
2001$-$3	&	20	38	59.29	&	42	22	48.8	&	4	\\
2001$-$4	&	20	38	59.32	&	42	22	49.6	&	5	\\
2001$-$5	&	20	38	59.35	&	42	22	47.4	&	6	\\
2001$-$6	&	20	38	59.56	&	42	23	5.3	&	9	\\
2001$-$7	&	20	38	59.64	&	42	22	47.2	&	10	\\
2001$-$8	&	20	38	59.72	&	42	22	49.6	&	15	\\
2001$-$9	&	20	38	59.73	&	42	23	16.2	&	12	\\
2001$-$10	&	20	38	59.73	&	42	22	47.6	&	14	\\
2001$-$11	&	20	38	59.76	&	42	22	45.5	&	16	\\
2001$-$12	&	20	38	59.83	&	42	22	50.1	&	17	\\
2001$-$13	&	20	38	59.89	&	42	22	46.0	&	18	\\
2001$-$14	&	20	38	59.92	&	42	22	47.9	&	22	\\
2001$-$15	&	20	38	59.92	&	42	22	50.3	&	21	\\
2001$-$16	&	20	38	59.93	&	42	22	45.0	&	20	\\
2001$-$17	&	20	38	59.99	&	42	22	35.2	&	23	\\
2001$-$18	&	20	39	0.08	&	42	22	47.1	&	24	\\
2001$-$19	&	20	39	0.09	&	42	22	44.7	&	25	\\
2001$-$20	&	20	39	0.12	&	42	22	43.7	&	26	\\
2001$-$21	&	20	39	0.13	&	42	22	47.1	&	28	\\
2001$-$22	&	20	39	0.14	&	42	22	48.8	&	27	\\
2001$-$23	&	20	39	0.19	&	42	22	47.7	&	29	\\
2001$-$24	&	20	39	0.26	&	42	22	46.0	&	30	\\
2001$-$25	&	20	39	0.29	&	42	22	47.4	&	32	\\
2001$-$26	&	20	39	0.33	&	42	22	48.5	&	34	\\
2001$-$27	&	20	39	0.49	&	42	22	46.3	&	35	\\
2001$-$28	&	20	39	0.55	&	42	22	47.7	&	36	\\
2001$-$29	&	20	39	0.60	&	42	22	45.0	&	37	\\
2001$-$30	&	20	39	1.05	&	42	22	41.6	&	40	\\
2001$-$31	&	20	39	1.05	&	42	22	17.6	&	38	\\
2001$-$32	&	20	39	1.09	&	42	22	1.0	&	41	\\
2001$-$33	&	20	39	1.17	&	42	22	6.8	&	42	\\
2001$-$34	&	20	39	1.18	&	42	22	6.8	&	42	\\
2001$-$35	&	20	39	1.22	&	42	22	41.2	&	43	\\
2001$-$36	&	20	39	1.47	&	42	23	6.3	&	44	\\
2001$-$37	&	20	39	1.48	&	42	22	7.6	&	45	\\
2001$-$38	&	20	39	1.50	&	42	22	45.2	&	46	\\
2001$-$39	&	20	39	1.51	&	42	22	40.8	&	47	\\
2001$-$40	&	20	39	1.54	&	42	22	2.8	&	None	\\
2001$-$41	&	20	39	1.67	&	42	22	43.4	&	48	\\
2001$-$42	&	20	39	1.70	&	42	22	45.2	&	49	\\
2001$-$43	&	20	39	1.76	&	42	22	45.3	&	50	\\
2001$-$44	&	20	39	2.02	&	42	22	41.2	&	52	\\
2001$-$45	&	20	39	2.02	&	42	22	40.8	&	51	\\
2001$-$46	&	20	39	2.08	&	42	22	45.2	&	53	\\
2001$-$47	&	20	39	2.21	&	42	22	43.4	&	54	\\
2001$-$48	&	20	39	2.22	&	42	22	42.6	&	55	\\
2001$-$49	&	20	39	2.24	&	42	22	44.0	&	56	\\
\enddata
\end{deluxetable}

A notable difference between the 2001 and 2012 data is in the locations of the strongest masers. Between 2001 and 2012, the two strongest masers in the field flipped in strength, so that the brightest maser in 2001 became the second brightest maser in 2012, and vice versa. In 2001, the strongest maser was located in the outer arc of the western lobe (currently at the location of maser 4 in Figure \ref{fig:2017smallmap}).  The second strongest maser was located about 10\arcsec\ away in the inner arc of the eastern lobe  (currently at the location of maser 30 in Figure \ref{fig:2017smallmap}). However, in 2012, the opposite is true: the strongest maser is located in the inner arc of the western lobe (currently at the location of maser 30 in Figure \ref{fig:2017smallmap}) while the second strongest maser is in the outer arc (currently at the location of maser 4 in Figure \ref{fig:2017smallmap}). This has remained unchanged from the 2012 to 2017 data; maser 30 in the inner arc is the strongest in the 2017 data, and maser 4 in the outer arc is the second strongest maser.

\subsection{Maser Variability} \label{sec:var}

The discussion in the previous section has established that whereas most of the masers have been stable entities over more than a 15-yr period between 2001 and 2017, there have likely been changes in the intensities of individual masers. Direct comparisons can be made between the 2012 and 2017 data, but such comparisons between the 2001 and 2017 data are difficult due to the difference in velocity resolution (0.66~\kms\ vs.\ 0.0266~\kms\ respectively). An examination of the qualitative trends (i.e., whether intensities have increased or decreased) between these epochs can be done by smoothing the higher velocity resolution data down %to that of the 2001 dataset. Henceforth, we will refer to this as our ``low velocity resolution comparison.''

%\clearpage

\onecolumngrid

\startlongtable
\begin{deluxetable}{ccccccc}
\tablenum{6}
\tablecaption{Observed Masers in \droh\ from M-S (\citealt{momjian+2017}) \label{tab:2012masers}}
\tablehead{
	\colhead{(1)} &  \colhead{(2)} &  \colhead{(3)} & \colhead{(4)} & \colhead{(5)} & \colhead{(6)}  & \colhead{(7)}   \\
 	\colhead{M-S} & \colhead{R.A.} & \colhead{Decl.} & \colhead{Intensity} & \colhead{Center Velocity} & \colhead{Velocity Linewidth} & \colhead{Corresponding} \\
	\colhead{Maser ID} & \colhead{(J2000)} & \colhead{(J2000)} & \colhead{(\jybeam)} & \colhead{(\kms)} & \colhead{(\kms)} & \colhead{2017 maser} }
\startdata
2012$-$1a	&	20	38	59.25	&	42	22	48.7	& 219.23 $\pm$ 3.91 & 0.826 $\pm$ 0.002 & 0.365 $\pm$ 0.002 & 4a \\
2012$-$1b	&	$\cdots$			&	$\cdots$		& $82.74 \pm 2.13$ & $ 0.531 \pm 0.009$ &   $0.484 \pm 0.010$ & 4b \\
2012$-$2	&	20	38	59.29	&	42	22	47.5	& $   7.11 \pm 0.12$ & $-0.855 \pm 0.003$ &   $0.372 \pm 0.007$ & 6a	\\
2012$-$3	&	20	38	59.31	&	42	22	49.6	& $  16.33 \pm 0.12$ & $ 1.099 \pm 0.001$ &   $0.352 \pm 0.003$ & 5	\\
2012$-$4	&	20	38	59.33	&	42	22	47.3	& $   6.64 \pm 0.10$ & $-1.004 \pm 0.004$ &   $0.300 \pm 0.008$ & 7 \\
2012$-$5	&	20	38	59.70	&	42	22	42.2	& $   0.35 \pm 0.01$ & $ 0.010 \pm 0.010$ &   $0.786 \pm 0.027$ & 11	\\
2012$-$6	&	20	38	59.71	&	42	22	45.8	& $   0.52 \pm 0.04$ & $-0.101 \pm 0.014$ &   $0.402 \pm 0.034$ & 13 \\
2012$-$7a	&	20	38	59.71	&	42	22	49.6	& $   1.03 \pm 0.34$ & $ 0.202 \pm 0.010$ &   $0.264 \pm 0.041$ & 15a	\\
2012$-$7b		&	$\cdots$			&	$\cdots$	& $   0.49 \pm 0.15$ & $-0.036 \pm 0.128$ &   $0.475 \pm 0.148$ & 15b	\\
2012$-$8	&	20	38	59.76	&	42	22	45.2	& $   0.72 \pm 0.04$ & $-0.249 \pm 0.010$ &   $0.365 \pm 0.025$ & 16		\\
2012$-$9	&	20	38	59.85	&	42	22	45.9	& $  14.79 \pm 0.09$ & $-0.216 \pm 0.001$ &   $0.219 \pm 0.003$ & 18a	\\
2012$-$10	&	20	38	59.85	&	42	22	46.2	& $   3.80 \pm 0.04$ & $-0.510 \pm 0.002$ &   $0.227 \pm 0.004$  & 18b		\\
2012$-$11	&	20	38	59.89	&	42	22	45.4	& $  16.46 \pm 0.12$ & $ 0.312 \pm 0.002$ &   $0.241 \pm 0.004$ & 19		\\
2012$-$12	&	20	38	59.91	&	42	22	45.0	& $  18.11 \pm 0.14$ & $ 0.561 \pm 0.001$ &   $0.203 \pm 0.001$ &	20\\
2012$-$13	&	20	38	59.96	&	42	22	35.2	& $   0.75 \pm 0.05$ & $-1.836 \pm 0.009$ &   $0.267 \pm 0.021$ & 23		\\
2012$-$14	&	20	39	0.15	&	42	22	47.8	& $  10.48 \pm 0.07$ & $-1.452 \pm 0.001$ &   $0.387 \pm 0.003$ &	29	\\
2012$-$15a	&	20	39	0.23	&	42	22	45.9	& $ 323.20 \pm 1.22$ & $ 0.396 \pm 0.001$ &   $0.309 \pm 0.002$ &	30a	\\
2012$-$15b	&	$\cdots$			&	$\cdots$	& $  44.30 \pm 1.22$ & $ 0.098 \pm 0.008$ &   $0.309 \pm 0.011$ &	30b	\\
2012$-$16	&	20	39	0.25	&	42	22	47.3	& $  11.79 \pm 0.10$ & $ 0.095 \pm 0.001$ &   $0.279 \pm 0.003$ &	32 	\\
2012$-$17a	&	20	39	0.25	&	42	22	46.5	& $   0.83 \pm 0.05$ & $ 1.240 \pm 0.016$ &   $0.605 \pm 0.041$ &	31b	\\
2012$-$17b	&	$\cdots$			&	$\cdots$	& $   7.99 \pm 0.06$ & $ 0.367 \pm 0.001$ &   $0.368 \pm 0.003$ & 31a	\\
2012$-$18	&	20	39	0.30	&	42	22	47.1	& $   0.90 \pm 0.01$ & $-2.049 \pm 0.003$ &   $0.312 \pm 0.007$ &	33a	\\
2012$-$19	&	20	39	0.33	&	42	22	48.3	& $   1.15 \pm 0.01$ & $-0.029 \pm 0.002$ &   $0.455 \pm 0.004$	&	34	\\
2012$-$20	&	20	39	0.51	&	42	22	47.6	& $   0.72 \pm 0.03$ & $ 1.352 \pm 0.009$ &   $0.270 \pm 0.018$ &	36c	\\
2012$-$21a	&	20	39	0.53	&	42	22	47.6	& $   1.82 \pm 0.04$ & $ 2.524 \pm 0.012$ &   $0.468 \pm 0.023$	&	36a	\\
2012$-$21b	&	$\cdots$			&	$\cdots$	& $   1.09 \pm 0.19$ & $ 2.116 \pm 0.025$ &   $0.300 \pm 0.054$ & 36b	\\
2012$-$22	&	20	39	0.52	&	42	22	47.6	& $   2.77 \pm 0.05$ & $ 1.783 \pm 0.009$ &   $0.367 \pm 0.022$	&	36d	\\
2012$-$23	&	20	39	1.01	&	42	22	41.6	& $   0.81 \pm 0.01$ & $-1.017 \pm 0.002$ &   $0.470 \pm 0.005$ &	39	\\
2012$-$24	&	20	39	1.20	&	42	22	41.0	& $   0.63 \pm 0.02$ & $-2.369 \pm 0.011$ &   $0.755 \pm 0.031$	&	43	\\
\enddata
\end{deluxetable}

\twocolumngrid

\noindent 
to that of the 2001 dataset. Henceforth, we will refer to this as our ``low velocity resolution comparison.'' \\ \\
Direct comparisons can be made between the 2012 and 2017 data, because they were obtained with similar angular resolution and the same velocity resolution. Henceforth, we will refer to this as our ``high velocity resolution comparison.'' To account for errors in the calibrator used to set the flux density scale of the source at this frequency, it is common practice to ignore any variations below the 10\% level. We adopt this standard, and report a variation only if it is larger than 10\%. Of the 24 masers listed in the 2012 data, 19 were fitted with one component, and 5 were fitted with two components. Of the 19 single-component masers in 2012, 15 varied in intensity (by more than the adopted 10\% threshold). Of the 5 masers fitted with two components in 2012, both components of two masers and one component of another showed variation. Out of these 20 maser spectral components that varied, only four increased in intensity from 2012 to 2017; they are maser 19, maser 33a, maser 30b, and maser 36b; maser numbers refer to the listing in Table~\ref{tab:2017masers}. The masers that decreased in intensity (again, by more than the 10\% threshold) from 2012 to 2017 are 5, 7, 11, 13, 16, 18a, 18b, 23, 30a, 31a, 32, 36a, 36c, 36d, 39, and 43. The two velocity components of maser 18 listed as 18a and 18b in Table~\ref{tab:2017masers} were registered in 2012 as two nearby but separate masers, and the four velocity components of maser~36 were registered in 2012 as three separate masers (Table~\ref{tab:2012masers}). Changes in the two strongest masers are discussed in more detail in Section~\ref{ss:m4}, and changes in several other masers in the western and eastern lobes are discussed in Section~\ref{ss:arcwest} and Section~\ref{ss:arceast} respectively.

\subsubsection{The two strongest masers: Maser 30 and Maser 4} \label{ss:m4}

The two strongest Class~I \meth\ masers at 44 GHz toward \droh\ flipped in rank from 2001 to 2012. Maser 30 in the inner arc of the western lobe was the brightest maser in the 2017 and 2012 data, whereas maser 4 was the second brightest (Table \ref{tab:2017masers} and Figure \ref{fig:2017smallmap}). However, maser 4 was brighter than maser 30 in 2001. In order to get a sense of how this variation came about, we used the AIPS task XSMTH to smooth the 2012 data from \citet{momjian+2017} to the same velocity resolution as the 2001 data. From this low velocity resolution comparison, we found that maser 4 became 1.6 times brighter from 2001 to 2012, but maser 30 became 7.7 times brighter during that same period. Thus maser 30 went through a much larger increase in intensity from 2001 to 2012 compared to maser 4. 

\vsp{0.1}

Next, both the 2012 and 2017 observations have similar angular resolution and the same velocity resolution. Therefore, direct comparisons can be made between these two epochs; we have labeled this the high velocity resolution comparison. Neither the stronger nor the weaker component of maser 4 varied above the 10\% level between 2012 and 2017. However, the stronger component of maser 30 showed a marginal decrease of 14\% from 2012 to 2017, whereas the weaker component showed a significant increase of 29\% from 44.30~\jybeam\ to 57.42~\jybeam. One interpretation of these numbers is that maser 4 in the outer arc of the western lobe has reached a quieter phase after 2012, whereas changes are still going on in maser 30 in the inner arc of the western lobe. 

\vsp{0.1}

\subsubsection{Masers in the western lobe} \label{ss:arcwest} 

In addition to the two strongest masers, there are indications of variability in several other masers in \droh. Of the 11 masers in the outer arc of the western lobe (Table~\ref{tab:ArcMasers}), maser 4 has already been discussed in Section~\ref{ss:m4} above. Of the other ten, maser 5 has intensity 13.65~\jybeam, masers 6, 7, 15, 34 and 36 have intensities 1-10~\jybeam, and the rest have intensity $<1$~\jybeam. Maser 5 is located to the north of maser 4, almost at the head of the bowshock delineated by the outer arc in the western lobe. Masers 5, 6, 15, 34, and 36 went up in intensity from 2001 to 2017 (Figure~\ref{fig:2017changesMap}); this trend was obtained from the low velocity resolution comparison of the 2001 and 2017 data. High velocity resolution comparison of the 2012 and 2017 data showed that practically all of these masers either decreased in intensity from 2012 to 2017, or remained unchanged at the 10\% level. For example, maser 5 decreased by 16.4\% from 16.33~\jybeam\ in 2012 to 13.65~\jybeam\ in 2017. The only exception was one of the four components of maser 36, which increased in intensity by 12.8\% from 1.09~\jybeam\ to 1.23~\jybeam, although only marginally above the 10\% level. Since masers 5, 6, 15, 34, and 36 went up in intensity from 2001 to 2017, but decreased or stayed the same from 2012 to 2017, they must have increased in intensity from 2001 to 2012. 

\vsp{0.1}

There are 20 masers in the inner arc of the western lobe (Table~\ref{tab:ArcMasers}). Of these, maser 30 has already been discussed in Section~\ref{ss:m4}. Of the rest, masers 18, 19, 20, 29, and 32 have intensities between 10-20~\jybeam, masers 25, 26, 28, 31, and 33 have intensities between 1-10~\jybeam, and the rest have intensities $<$ 1~\jybeam. Masers 11, 13, 31, and 33 were not present in 2001 but were observed in 2012, and masers 19 and 20 may have been reported as one maser in 2001. For the 15 masers that were present both in 2001 and 2017 (including maser 30), the intensity increased for all but three masers (Figure~\ref{fig:2017changesMap}). The three exceptions are maser 24 which went down in intensity from 2001 to 2017, and masers 14 and 37 which showed no changes above the 10\% level. These trends were obtained from our low velocity resolution comparison of the 2001 and 2017 data. Masers 16, 18, 29, 30, and 32 were present in all three epochs, 2001, 2012, and 2017; masers 19 and 20 were present as separate masers in 2012 and 2017, but may have been observed as one maser in 2001. High velocity resolution comparison of the 2012 and 2017 data showed that practically all of these masers, including the four listed above that were observed in 2012 but not in 2001, either decreased in intensity from 2012 to 2017, or remained unchanged at the 10\% level. For example, the stronger component of maser 31 decreased to less than half its intensity, from 7.99~\jybeam\ in 2012 to 3.59~\jybeam\ in 2017. The only two exceptions are the weaker component of maser 30 which increased by 29.6\%, and maser 33a which increased from 0.9~\jybeam\ to 1.23~\jybeam. 

In summary, masers in both the inner and outer arc of the western lobe display the same general trend. Most, with only a few exceptions, appear to have increased in intensity from 2001 to 2012, then gone down or stayed the same from 2012 to 2017. In particular, increases in the inner arc appear to be more than in the outer arc. This is discussed in more detail in Section~\ref{sec:reasons}, along with its implications.

\subsubsection{Masers in the eastern lobe} \label{ss:arceast} 

There are 5 masers in the inner arc of the eastern lobe (Table~\ref{tab:ArcMasers}). Our low velocity resolution comparison of 2017 and 2001 data shows that masers 40 and 43 have decreased in intensity from 2001 to 2017, whereas masers 46 and 48 have increased in intensity (Figure~\ref{fig:2017changesMap}). Maser 39 was not observed in 2001. Masers 39, 40, and 43 are redshifted masers, even though they are present in the eastern lobe. Although maser 40 was below the 0.3~\jybeam\ detection limit imposed for the 2012 observations, masers 39 and 43 are present in the 2012 data. Our high velocity resolution comparison reveals that maser 39 decreased by 18.5\% from 0.81~\jybeam\ in 2012 to 0.66~\jybeam\ in 2017. Maser 43 decreased from 0.63~\jybeam\ to 0.56~\jybeam, just marginally above the 10\% limit. With a center velocity of $-$2.83~\kms, maser 46 is also redshifted, but the velocity would have been just outside the most redshifted channel imaged in 2012. Maser 48 is blueshifted and would not have been detected in 2012.

\vsp{0.1}

There are 10 masers in the outer arc of the eastern lobe (Table~\ref{tab:ArcMasers}). All of these masers are blueshifted, so they were not observed in 2012. Other than one (maser 57), all of them were also observed in 2001. Our low velocity resolution comparison showed that seven of these masers changed in intensity, and two did not change. Of the seven, three increased and four decreased in intensity (Figure~\ref{fig:2017changesMap}). Since no data are available from 2012 for these masers, it is not possible to say if this increase has been consistent from 2001 to 2017, or whether an increase from 2001 until 2012 was followed by a decrease, similar to the masers in the western lobe.

\vsp{0.1}

\subsection{Reasons for maser variability} \label{sec:reasons}

The variability of 44 GHz Class I \meth\ masers has been hinted at in several papers \citep{leurini+2016, kurtz+2004, momjian+2012}. In this paper, we have presented a dedicated investigation of such masers in the high mass star forming region \droh\ and established that such masers vary over the long term. A theoretical model for why such variability should occur is beyond the scope of this paper. Nevertheless, some of the processes that could cause variability in Class I \meth\ masers are discussed qualitatively in this section. 

\vsp{0.1}

Masers can vary due to changes in pumping or due to changes in the path length over which the signal is amplified. A change in pumping could be caused by the propagation of a shock front through the region. It is unlikely, however, that the same shock front that caused the increase in maser 30 in the inner arc of the western lobe from 2001 to 2012 has led to the increase in maser 4 in the outer arc. The distance between maser 30 and maser 4 is about 10\arcsec, equivalent to 0.07~pc. To travel from the location of maser 30 to the location of maser 4 in the period from 2001 to 2012 would require the shock front to be traveling at an absurdly high velocity over 5000~\kms. Instead, it is likely that we are seeing an instance of episodic accretion.

\vsp{0.1}

Episodes of increased accretion are well known in low mass star formation, but outbursts related to such increases in accretion have only recently been found in a few high mass star forming regions (e.g., \citealt{caratti+2017}; \citealt{brogan+2019}). Each episode of accretion is accompanied by an ejection event which registers as an outward-propagating shock (\citealt{caratti+2015}). Such a model would imply that the increase in intensity in the masers in the outer arc of the western lobe in \droh\ was caused by an earlier shock related to an enhanced accretion episode. This continued to cause a modest increase in the intensity of the masers through 2012, following which they have decreased or remained largely unchanged at the 10\% level. The increase in the intensity of masers in the inner arc of the western lobe could then be ascribed to a shock front created by a more recent accretion event that reached the inner arc sometime between 2001 and 2012. 

\vsp{0.1}

Activity in the inner arc is significantly greater than in the outer arc in the western lobe, suggesting that this model of episodic accretion outbursts is at least plausible. For example, the increase in maser intensities in the inner arc is larger than in the outer arc. In Section~\ref{ss:m4}, we have seen that maser 30 in the inner arc increased by 7.7 times from 2001 to 2012, whereas maser 4 in the outer arc increased only by a 1.6 factor. Just like the increases in the two strongest masers, masers with intensities between 1-10~\jybeam\ in the outer arc increased by 3-5 times, whereas those in the inner arc increased by 3-8 times. Moreover, only one new low intensity maser showed up in 2017 in the outer arc of the western lobe that was not present in 2001; maser 8 has intensity 0.72~\jybeam\ (Table~\ref{tab:2017masers}). Meanwhile, four new masers showed up in the inner arc of the western lobe in 2017 (and 2012) that were not present in 2001; masers 11 and 13 are low intensity, but maser~31a has an intensity of 3.59~\jybeam, and maser~33a is 1.23~\jybeam\ in the 2017 observations (Table~\ref{tab:2017masers}). The effects of such episodic accretion aren't as clear in the eastern lobe.

\vsp{0.1}

Additional support for the model of outbursts related to accretion comes from considering the time between such episodes. If we assume that methanol masers are caused by slower-moving shocks (\citealt{leurini+2016}) and use a shock velocity of 20~\kms, then it would take about 3500~yr for a shock to propagate from the position of maser 30 in the inner arc to maser 4 in the outer arc. Therefore, the two accretion events that led to maser variation in the outer arc and the inner arc of the western lobe would have occurred at an interval of about 3500~yr. If we use a higher shock velocity of 30~\kms, then the accretion events would have occurred at an interval of 2300~yr. In their numerical simulation of accretion-driven bursts in high mass star formation, \citet{meyer+2017} found such accretion bursts to occur at intervals of 3000~yr, consistent with the calculation for \droh\ above. Hence, if the shock that caused the increase in intensity of maser 30 and other masers in the inner arc reached the inner arc of the western lobe sometime after 2001, the shock that caused the increase in intensity of maser 4 and other masers in the outer arc would have reached it about 2300-3500~yr earlier (depending on how fast the shock was traveling). The modest increase in maser 4 after 2001 may be a lingering effect of this earlier shock, or it could be a result of the alternative scenarios discussed below.

\vsp{0.1}

Additional scenarios that can also cause variability in masers arise from changes in the path length over which the maser signal is amplified. One possibility is to assume that the maser region has an ellipsoidal shape and spins as a solid body with constant angular momentum so that there is velocity coherence along the line of sight, like \citet{andreev+2017} did to model the variation of formaldehyde (H$_2$CO) masers. As the maser region rotates (slowly), the amplification path changes and causes variations in the intensity of the maser. Another possibility is to assume a foreground cloud moving in front of a maser cloud, as \citet{boboltz+1998} did to model variation in H$_2$O masers in the high mass star forming region W49. Yet another cause of maser variability could be through an increase in the path length due to multiple overlapping maser clouds along the line of sight (\citealt{burns+2020}). All of these, a rotating maser cloud, a foreground cloud moving in front of a maser cloud, and overlapping maser clouds, could be initiated by turbulence, which can be caused by the very same shocks that are responsible for pumping Class~I \meth\ masers in outflows. 

\vsp{0.1}

\section{Conclusion} \label{sec:conc}

We observed 57 Class~I \meth\ maser spots at 44 GHz toward the high mass star forming region \droh\ in May 2017. The masers are arranged in a western and eastern lobe with an inner and an outer arc in each lobe, consistent with previous observations. The center velocities of the masers range from $-8.65$~\kms\ to +2.56~\kms. 

\vsp{0.1}

In order to investigate the variability of 44 GHz Class I \meth\ masers in \droh, we compared our 2017 results to observations made in 2012 \citep{momjian+2017} and observations in 2001 taken from the literature \citep{araya+2009}. This is likely  the first dedicated study of the variability of Class I \meth\ masers in high mass star forming regions. Knowing the variability of such masers allows us to better understand the maser phenomenon, and such an understanding makes these masers a more effective probe of the high mass star formation process. Much remains to be learned about the process of high mass star formation, because high mass stars are rare and located farther away from us. Masers are bright and compact sources and serve as excellent probes of high mass star forming regions at high angular resolution.

\vsp{0.1}

The 57 masers in our 2017 data include all the 24 redshifted masers observed in 2012 \citep{momjian+2017}, and 48 of the 49 masers observed in 2001 \citep{araya+2009}. With a few exceptions, most of the masers in both the inner and the outer arcs in the western lobe increased in intensity from 2001 to 2012, then decreased from 2012 to 2017, or stayed constant. Most masers in the outer arc of the eastern lobe also changed in intensity. For the four masers in the inner arc of the eastern lobe for which data are available to compare, two out of four masers increased in intensity, whereas the other two decreased.

\vsp{0.1}

In particular, there is indication of significant variation in intensity for the two strongest masers in \droh. From 2001 to 2012, maser 4 in the outer arc of the western lobe increased in intensity by a factor of 1.6, whereas maser 30 in the inner arc increased by 7.7 times. As a result, maser 4 became the second brightest maser in \droh; it had been the brightest maser in 2001. Also as a result, maser 30 rose to become the brightest; it had been the second brightest maser in 2001. This, along with changes in intensity in several other masers, leads us to conclude with certainty that 44 GHz Class I \meth\ masers do exhibit variability on long timescales of 5-10~yr. 

\vsp{0.1}

Although a theoretical model for the variability of Class~I~\meth\ masers is beyond the scope of this paper, we have presented plausible scenarios that could cause such variability. One is that of episodic accretion, which is now known to occur in high mass star formation. Each such accretion episode is accompanied by an ejection event, which then creates the bowshocks in outflows such as those observed in \droh. If the variability of the Class~I \meth\ masers in the inner and outer arcs of the western lobe of \droh\ has been caused by such an event, typical shock velocities would constrain these accretion events to be spread apart by about 3500~yr, which is consistent with the results of numerical simulations in the literature. There is certainly more activity going on in the inner arc of the western lobe compared to the outer arc to support this picture of a more recent shock to the inner arc that is causing these changes. Examples of such activity include larger increases in intensity and more new masers in the inner arc. Alternatively, changes in maser intensities could be a result of turbulence in the shocked regions. Such turbulence could cause ellipsoidal maser regions to rotate, or a foreground cloud to come in front of a maser region, or bring multiple overlapping maser clouds into alignment along the line of sight. All of these would cause a change in the amplification path length and lead to variations in maser intensity. 

\vsp{0.1}

Much remains to be done regarding the variability of Class~I \meth\ masers. Future work should focus on two fronts. The first would consist of establishing the variability of Class~I \meth\ masers in a larger number of regions. The second front would entail looking for the causes of such variability. Extending to a larger number of regions would demonstrate how prevalent is the phenomenon of variability in Class~I \meth\ masers over the longer term. Looking for the causes of such variability would involve looking at other maser lines to see if they varied on similar timescales, and if the accretion event was recorded in some other tracer.

\vsp{0.5}

\acknowledgments
We thank an anonymous referee for suggestions that have significantly enhanced the content and improved the presentation of material in this paper. NW gratefully acknowledges a Summer Student Research Assistantship %in 2020 
from the NRAO. The results reported in this paper formed part of the Master's thesis of NW at DePaul University.

%\vspace{5mm}
\facilities{VLA}

\clearpage

\clearpage

\bigskip

\clearpage

\clearpage

\clearpage

%FIGURES

\clearpage


\begin{thebibliography}{}

\bibitem[Andreev et al.(2017)]{andreev+2017} Andreev, N., Araya, E.~D., Hoffman, I.~M., et al.\ 2017, ApJS, 232, 29. 
\bibitem[Araya et al.(2009)]{araya+2009} Araya, E.~D., Kurtz, S., Hofner, P., et al.\ 2009, \apj, 698, 1321.
\bibitem[Arce et al.(2007)]{arce+2007} Arce, H.~G., Shepherd, D., Gueth, F., et al.\ 2007, Protostars and Planets V, 245
\bibitem[Boboltz et al.(1998)]{boboltz+1998} Boboltz, D.~A., Simonetti, J.~H., Dennison, B., et al.\ 1998, ApJ, 509, 256. % APS July 9
\bibitem[Brogan et al.(2019)]{brogan+2019} Brogan, C.~L., Hunter, T.~R., Towner, A.~P.~M., et al.\ 2019, ApJL, 881, L39. 
\bibitem[Burns et al.(2020)]{burns+2020} Burns, R.~A., Orosz, G., Bayandina, O., et al.\ 2020, \mnras, 491, 4069. % added post-refereed version
\bibitem[Caratti o Garatti et al.(2015)]{caratti+2015} Caratti o Garatti, A., Stecklum, B., Linz, H., et al.\ 2015, A\&A, 573, A82. 
\bibitem[Caratti o Garatti et al.(2017)]{caratti+2017} Caratti o Garatti, A., Stecklum, B., Garcia Lopez, R., et al.\ 2017, Nature Physics, 13, 276. % APS July 9
\bibitem[Chandler et al.(1993)]{chandler+1993} Chandler, C.~J., Moore, T.~J.~T., Mountain, C.~M., et al.\ 1993, MNRAS, 261, 694. 
\bibitem[Girart et al.(2013)]{girart+2013} Girart, J.~M., Frau, P., Zhang, Q., et al.\ 2013, \apj, 772, 69. doi:10.1088/0004-637X/772/1/69
\bibitem[Hunter et al.(2018)]{hunter+2018} Hunter, T.~R., Brogan, C.~L., MacLeod, G.~C., et al.\ 2018, \apj, 854, 170 % added APS July 7
\bibitem[Kogan \& Slysh(1998)]{kogan+1998} Kogan, L. \& Slysh, V.\ 1998, \apj, 497, 800.
\bibitem[Kurtz et al.(2004)]{kurtz+2004} Kurtz, S., Hofner, P., \& Alvarez, C.~V.\ 2004, ApJS, 155, 149.
\bibitem[Leurini et al.(2016)]{leurini+2016} Leurini, S., Menten, K.~M., \& Walmsley, C.~M.\ 2016, \aap, 592, A31. 
\bibitem[Meyer et al.(2017)]{meyer+2017} Meyer, D.~M.-A., Vorobyov, E.~I., Kuiper, R., et al.\ 2017, MNRAS, 464, L90.
\bibitem[Momjian \& Sarma(2012)]{momjian+2012} Momjian, E. \& Sarma, A.~P.\ 2012, \aj, 144, 189.
\bibitem[Momjian \& Sarma(2017)]{momjian+2017} Momjian, E. \& Sarma, A.~P.\ 2017, \apj, 834, 168.
\bibitem[Motte et al.(2018)]{motte+2018} Motte, F., Bontemps, S., \& Louvet, F.\ 2018, \araa, 56, 41. 
\bibitem[Orozco-Aguilera et al.(2019)]{orozco+2019} Orozco-Aguilera, M.~T., Hern{\'a}ndez-G{\'o}mez, A., \& Zapata, L.~A.\ 2019, \aj, 157, 20. % added to post-refereed version
\bibitem[Richards et al.(2020)]{richards+2020} Richards, A.~M.~S., Sobolev, A., Baudry, A., et al.\ 2020, Advances in Space Research, 65, 780.
\bibitem[Rygl et al.(2012)]{rygl+2012} Rygl, K.~L.~J., et al.\ 2012, \aap, 539, A79.% Brunthaler, A., Sanna, A., et al.\ 2012, \aap, 539, A79.
\bibitem[Zapata et al.(2012)]{zapata+2012} Zapata, L.~A., Loinard, L., Su, Y.-N., et al.\ 2012, \apj, 744, 86.
\bibitem[Zinnecker \& Yorke(2007)]{zy+2007} Zinnecker, H. \& Yorke, H.~W.\ 2007, \araa, 45, 481. 

\end{thebibliography}
\end{document}